\documentclass[aps,pra,floatfix,twocolumn,superscriptaddress,showpacs,10pt]{revtex4-1}

\usepackage[latin9]{inputenc}
\usepackage{amsmath}
\usepackage{amssymb}
\usepackage{graphicx}
\usepackage{esint}
\usepackage{times}
\usepackage{array}

\usepackage[unicode=true,
bookmarks=true,bookmarksnumbered=false,bookmarksopen=false,
breaklinks=false,pdfborder={0 0 1},backref=false,colorlinks=true]
{hyperref}
\hypersetup{
    linkcolor=magenta, urlcolor=blue, citecolor=blue, pdfstartview={FitH}, hyperfootnotes=false, unicode=true}

\makeatletter

\pdfpageheight\paperheight
\pdfpagewidth\paperwidth

\@ifundefined{textcolor}{}
{%
    \definecolor{BLACK}{gray}{0}
    \definecolor{WHITE}{gray}{1}
    \definecolor{RED}{rgb}{1,0,0}
    \definecolor{GREEN}{rgb}{0,1,0}
    \definecolor{BLUE}{rgb}{0,0,1}
    \definecolor{CYAN}{cmyk}{1,0,0,0}
    \definecolor{MAGENTA}{cmyk}{0,1,0,0}
    \definecolor{YELLOW}{cmyk}{0,0,1,0}
}

\usepackage{xcolor}\usepackage{soul}

\newcommand{\bra}[1]{\ensuremath{\left\langle#1\right|}}
\newcommand{\ket}[1]{\ensuremath{\left|#1\right\rangle}}
\definecolor{blue}{rgb}{0,0,1}
\definecolor{red}{rgb}{1,0,0}
\definecolor{green}{rgb}{0,1,0}

\makeatother

\begin{document}
\title{Superrobust Geometric Control of a Superconducting Circuit}

\author{Sai Li}
\thanks{These authors contributed equally to this work.}
\affiliation{Shenzhen Institute for Quantum Science and Engineering, Southern University of Science and Technology, Shenzhen, Guangdong 518055, China}
\affiliation{International Quantum Academy, Futian District, Shenzhen, Guangdong 518048, China}
\affiliation{Guangdong Provincial Key Laboratory of Quantum Science and Engineering, Southern University of Science and Technology, Shenzhen, Guangdong 518055, China}
\affiliation{Shenzhen Key Laboratory of Quantum Science and Engineering, Southern University of Science and Technology, Shenzhen, Guangdong 518055, China}

\author{Bao-Jie Liu}
\thanks{These authors contributed equally to this work.}
\affiliation{Shenzhen Institute for Quantum Science and Engineering, Southern University of Science and Technology, Shenzhen, Guangdong 518055, China}
\affiliation{International Quantum Academy, Futian District, Shenzhen, Guangdong 518048, China}
\affiliation{Department of Physics, Southern University of Science and Technology, Shenzhen, Guangdong 518055, China}
\affiliation{Guangdong Provincial Key Laboratory of Quantum Science and Engineering, Southern University of Science and Technology, Shenzhen, Guangdong 518055, China}
\affiliation{Shenzhen Key Laboratory of Quantum Science and Engineering, Southern University of Science and Technology, Shenzhen, Guangdong 518055, China}

\author{Zhongchu Ni}
\affiliation{Shenzhen Institute for Quantum Science and Engineering, Southern University of Science and Technology, Shenzhen, Guangdong 518055, China}
\affiliation{International Quantum Academy, Futian District, Shenzhen, Guangdong 518048, China}
\affiliation{Department of Physics, Southern University of Science and Technology, Shenzhen, Guangdong 518055, China}
\affiliation{Guangdong Provincial Key Laboratory of Quantum Science and Engineering, Southern University of Science and Technology, Shenzhen, Guangdong 518055, China}
\affiliation{Shenzhen Key Laboratory of Quantum Science and Engineering, Southern University of Science and Technology, Shenzhen, Guangdong 518055, China}

\author{Libo Zhang}
\affiliation{Shenzhen Institute for Quantum Science and Engineering, Southern University of Science and Technology, Shenzhen, Guangdong 518055, China}
\affiliation{International Quantum Academy, Futian District, Shenzhen, Guangdong 518048, China}
\affiliation{Guangdong Provincial Key Laboratory of Quantum Science and Engineering, Southern University of Science and Technology, Shenzhen, Guangdong 518055, China}
\affiliation{Shenzhen Key Laboratory of Quantum Science and Engineering, Southern University of Science and Technology, Shenzhen, Guangdong 518055, China}

\author{Zheng-Yuan Xue}
\affiliation{Guangdong Provincial Key Laboratory of Quantum Engineering and Quantum Materials, \\and School of Physics and Telecommunication Engineering, South China Normal University, Guangzhou 510006, China}
\affiliation{Guangdong-Hong Kong Joint Laboratory of Quantum Matter, and Frontier Research Institute for Physics, \\South China Normal University, Guangzhou 510006, China}
\affiliation{Guangdong Provincial Key Laboratory of Quantum Science and Engineering, Southern University of Science and Technology, Shenzhen, Guangdong 518055, China}

\author{Jian Li}
\affiliation{Shenzhen Institute for Quantum Science and Engineering, Southern University of Science and Technology, Shenzhen, Guangdong 518055, China}
\affiliation{International Quantum Academy, Futian District, Shenzhen, Guangdong 518048, China}
\affiliation{Guangdong Provincial Key Laboratory of Quantum Science and Engineering, Southern University of Science and Technology, Shenzhen, Guangdong 518055, China}
\affiliation{Shenzhen Key Laboratory of Quantum Science and Engineering, Southern University of Science and Technology, Shenzhen, Guangdong 518055, China}

\author{Fei Yan}
\affiliation{Shenzhen Institute for Quantum Science and Engineering, Southern University of Science and Technology, Shenzhen, Guangdong 518055, China}
\affiliation{International Quantum Academy, Futian District, Shenzhen, Guangdong 518048, China}
\affiliation{Guangdong Provincial Key Laboratory of Quantum Science and Engineering, Southern University of Science and Technology, Shenzhen, Guangdong 518055, China}
\affiliation{Shenzhen Key Laboratory of Quantum Science and Engineering, Southern University of Science and Technology, Shenzhen, Guangdong 518055, China}

\author{Yuanzhen Chen}
\affiliation{Shenzhen Institute for Quantum Science and Engineering, Southern University of Science and Technology, Shenzhen, Guangdong 518055, China}
\affiliation{International Quantum Academy, Futian District, Shenzhen, Guangdong 518048, China}
\affiliation{Department of Physics, Southern University of Science and Technology, Shenzhen, Guangdong 518055, China}
\affiliation{Guangdong Provincial Key Laboratory of Quantum Science and Engineering, Southern University of Science and Technology, Shenzhen, Guangdong 518055, China}
\affiliation{Shenzhen Key Laboratory of Quantum Science and Engineering, Southern University of Science and Technology, Shenzhen, Guangdong 518055, China}

\author{Song Liu}
\email{lius3@sustech.edu.cn}
\affiliation{Shenzhen Institute for Quantum Science and Engineering, Southern University of Science and Technology, Shenzhen, Guangdong 518055, China}
\affiliation{International Quantum Academy, Futian District, Shenzhen, Guangdong 518048, China}
\affiliation{Guangdong Provincial Key Laboratory of Quantum Science and Engineering, Southern University of Science and Technology, Shenzhen, Guangdong 518055, China}
\affiliation{Shenzhen Key Laboratory of Quantum Science and Engineering, Southern University of Science and Technology, Shenzhen, Guangdong 518055, China}

\author{Man-Hong Yung}
\email{yung@sustech.edu.cn}
\affiliation{Shenzhen Institute for Quantum Science and Engineering, Southern University of Science and Technology, Shenzhen, Guangdong 518055, China}
\affiliation{International Quantum Academy, Futian District, Shenzhen, Guangdong 518048, China}
\affiliation{Department of Physics, Southern University of Science and Technology, Shenzhen, Guangdong 518055, China}
\affiliation{Guangdong Provincial Key Laboratory of Quantum Science and Engineering, Southern University of Science and Technology, Shenzhen, Guangdong 518055, China}
\affiliation{Shenzhen Key Laboratory of Quantum Science and Engineering, Southern University of Science and Technology, Shenzhen, Guangdong 518055, China}

\author{Yuan Xu}
\email{xuy5@sustech.edu.cn}
\affiliation{Shenzhen Institute for Quantum Science and Engineering, Southern University of Science and Technology, Shenzhen, Guangdong 518055, China}
\affiliation{International Quantum Academy, Futian District, Shenzhen, Guangdong 518048, China}
\affiliation{Guangdong Provincial Key Laboratory of Quantum Science and Engineering, Southern University of Science and Technology, Shenzhen, Guangdong 518055, China}
\affiliation{Shenzhen Key Laboratory of Quantum Science and Engineering, Southern University of Science and Technology, Shenzhen, Guangdong 518055, China}

\author{Dapeng Yu}
\affiliation{Shenzhen Institute for Quantum Science and Engineering, Southern University of Science and Technology, Shenzhen, Guangdong 518055, China}
\affiliation{International Quantum Academy, Futian District, Shenzhen, Guangdong 518048, China}
\affiliation{Department of Physics, Southern University of Science and Technology, Shenzhen, Guangdong 518055, China}
\affiliation{Guangdong Provincial Key Laboratory of Quantum Science and Engineering, Southern University of Science and Technology, Shenzhen, Guangdong 518055, China}
\affiliation{Shenzhen Key Laboratory of Quantum Science and Engineering, Southern University of Science and Technology, Shenzhen, Guangdong 518055, China}


\begin{abstract}
Geometric phases accompanying adiabatic quantum evolutions can be used to construct robust quantum control for quantum information processing due to their noise-resilient feature. A significant development along this line is to construct geometric gates using nonadiabatic quantum evolutions to reduce errors due to decoherence. However, it has been shown that nonadiabatic geometric gates are not necessarily more robust than dynamical ones, in contrast to an intuitive expectation. Here we experimentally investigate this issue for the case of nonadiabatic holonomic quantum computation~(NHQC) and show that conventional NHQC schemes cannot guarantee the expected robustness due to a cross coupling to the states outside the computational space. We implement a different set of constraints for gate construction in order to suppress such cross coupling to achieve an enhanced robustness. Using a superconducting quantum circuit, we demonstrate high-fidelity holonomic gates whose infidelity against quasi-static transverse errors can be suppressed up to the fourth order, instead of the second order in conventional NHQC and dynamical gates. In addition, we explicitly measure the accumulated dynamical phase due to the above mentioned cross coupling and verify that it is indeed much reduced in our NHQC scheme. We further demonstrate a protocol for constructing two-qubit NHQC gates also with an enhanced robustness. 
\end{abstract}
\maketitle
\vskip 0.5cm

\narrowtext

\section{INTRODUCTION}
Robust quantum operations are essential for noisy intermediate-scale quantum computation~\cite{preskill2018quantum} with the existence of various error sources. Different strategies have been proposed for realizing such robust operations. One seminal example is quantum control based on geometric phases~\cite{sjoqvist2015geometric}. As a general and fundamental feature that accompanies quantum evolution, geometric phases are solely determined by global properties, rather than local details, of the evolution. Therefore, they are intrinsically robust against certain types of noise and control imperfections. Such a property has naturally been developed into the framework of geometric or holonomic quantum computation~\cite{Zanardi1999, sjoqvist2008trend}, where quantum gates based on abelian or nonabelian geometric phases are realized through carefully engineering the involved quantum evolutions~\cite{berry1984quantal, Aharonov1987Phase, Anandan1988, Zhu2005, Berger2013, Chiara2003, Leek2007, Filipp2009, Xu2020Experimental}. 

Early proposals of holonomic quantum computation utilize adiabatic evolution to suppress unwanted transitions among the instantaneous eigenstates of the Hamiltonians~\cite{Jones2000, duan2001geometric, Wu2005Holonomic}, which makes the resulting quantum gates have a long runtime and thus be sensitive to decoherence. To overcome such a problem, nonadiabatic holonomic quantum computation (NHQC)~\cite{Sjoqvist2012a, Xu2012Nonadiabatic} has been proposed to reduce the runtime of quantum gates~\cite{Xue2015,Herterich2016, Xue2017, Zhou2018, Hong2018, Mousolou2017,Liu2019,Zhao2020}. Various NHQC schemes have been experimentally demonstrated on different physical platforms, including superconducting circuits~\cite{Abdumalikov2013, Danilin2018, Egger2019, Xu2018, Han2020, Zhang2019,Yan2019,Optica21}, nuclear magnetic resonance~\cite{Feng2013, long2017, Zhu2019}, and nitrogen-vacancy centers in diamond~\cite{Zu2014,Arroyo2014, Sekiguchi2017, Zhou2017, Ishida2018,Nagata2018}. However, it has been found that such NHQC gates are not significantly more robust than the standard dynamical ones. For example, infidelity of both types of gates exhibits a second-order dependence of control errors~\cite{Johansson2012, Zheng2016, Ramberg2019, Jun2017, Xu2017,Li2020}. The missing of the ``intrinsic" robustness theoretically expected for the NHQC gates (in fact, for other types of geometric gates as well) remains a puzzle in the community and needs to be resolved before such gates can ever become practically useful.

Here, by reexamining the design principles of NHQC gates following a recent theoretical work~\cite{Liu2020}, we show that in conventional NHQC schemes, the phases used for gate construction may become a mixture of geometric and dynamical components due to a cross-coupling to the states outside the computational space, which compromises the robustness of geometric gates. This issue can be resolved by imposing a different set of constraints to the gate construction. Using a superconducting quantum circuit~\cite{Clarke2008, You2011, Devoret2013}, we have experimentally demonstrated arbitrary single-qubit NHQC gates complying with such constraints, with an average fidelity of 0.9956 characterized by the standard randomized benchmarking (RB). We have also measured explicitly the accumulated dynamical phase due to the cross-coupling discussed above and found that using our scheme, this phase is indeed much suppressed compared to conventional NHQC schemes. As a consequence, the dependence of gate infidelity on control errors (or quasistatic transverse noise) in our scheme is suppressed to the fourth order, achieving an enhanced robustness. We thus label our scheme as super robust NHQC (SR NHQC). Lastly, we have also demonstrated a protocol for constructing two-qubit SR-NHQC gates as well as their enhanced robustness.

\begin{figure}[t]
	\includegraphics{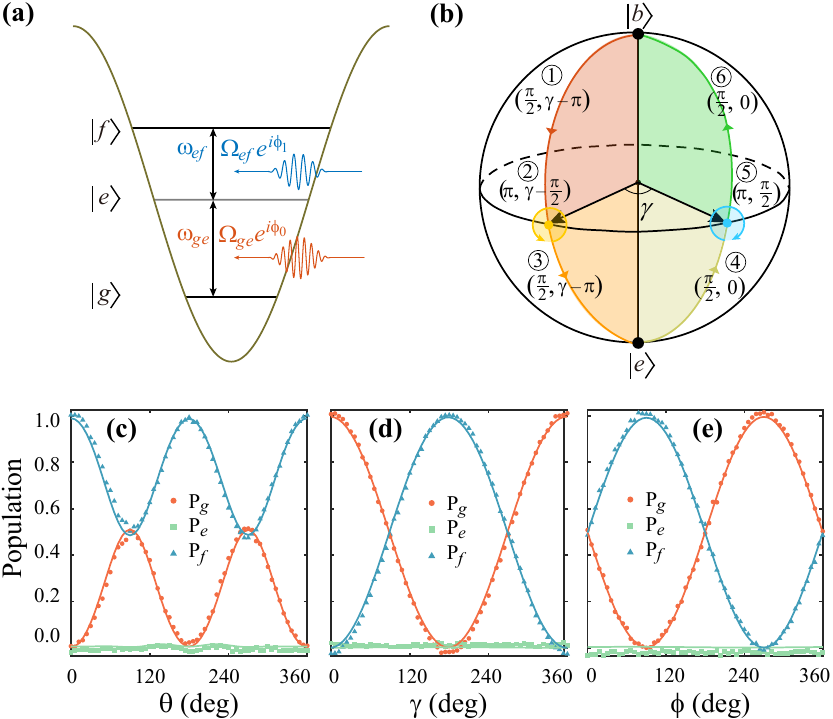}
	\caption{Super robust holonomic gates.
	(a) The lowest three energy levels of a superconducting transmon qubit driven by two resonant microwave pulses on the transitions of $\ket{g} \leftrightarrow \ket{e}$ and $\ket{e} \leftrightarrow \ket{f}$, respectively. 
      (b) Bloch sphere representation of quantum evolutions in the \{$\ket{b}$, $\ket{e}$\} subspace for the single-qubit SR-NHQC gate, which comprises six sequential rotations $R_{\varphi}(\vartheta)$ in the equatorial plane, whose rotation angles and axes $(\vartheta_{i}, \varphi_{i})$ $(i = 1,...,6)$ are specified alongside. 
    (c)-(e) The measured populations of the transmon qubit for single-qubit SR-NHQC gates $U_1(\theta, 0, \pi/2)$, $U_1(\pi/2,0,\gamma)$, and $U_1(\pi/2,\phi,\pi/2)$. The initial states in the three cases are $\ket{f}$, $\ket{g}$, and $\left(\ket{g}+\ket{f}\right)/\sqrt{2}$, respectively. Symbols and solid lines are experimental data and numerical simulations.}
    \label{fig1}
\end{figure}

\section{SR-NHQC SCHEME}
We first present the theoretical framework of the SR-NHQC scheme for a single-qubit case. Consider a typical NHQC scheme comprising three states of $\{\ket{g},\ket{f},\ket{e}\}$ driven by two resonant pulses with time-dependent amplitudes $\Omega_{ge}(t)$ and $\Omega_{ef}(t)$, and phases $\phi_0 (t)$ and $\phi_1 (t)$ (Fig.~\ref{fig1}(a)). Here $\ket{g}$ and $\ket{f}$ form the computational basis and $\ket{e}$ is an ancillary state. Under the rotating-wave approximation, the system Hamiltonian reads (with $\hbar \equiv1$)~\cite{Li2011} 
\begin{equation}
H(t)=\frac{1}{2}\left[\Omega_{ge}(t) e^{i \phi_{0}(t)}|g\rangle\langle e|+\Omega_{ef}(t) e^{i \phi_{1}(t)}| f\rangle\langle e|\right]+\mathrm{H.c.}\notag
\end{equation}
By defining a bright state as $| b\rangle \equiv -\sin(\frac{\theta}{2})e^{-i\phi }| g\rangle +\cos(\frac{\theta}{2})|f\rangle$ where $\phi \equiv \phi_1(t)-\phi_0(t)-\pi$ and $\tan(\theta/2) \equiv \Omega_{ge}(t)/\Omega_{ef}(t)$, one has $H(t)=\frac{1}{2}\left[\Omega(t)e^{i\phi_{1}(t)}| b\rangle \langle e| +\mathrm{H.c.}\right]$ with $\Omega(t) \equiv \sqrt{\Omega_{ge}(t)^2+\Omega_{ef}(t)^2}$. We further define a dark state $|d\rangle=\cos (\theta/2)e^{-i \phi}|g\rangle+\sin (\theta/2) |f\rangle$, which is decoupled from the system evolution since $H(t)|d\rangle=0$. In the following discussion, we keep both $\theta$ and $\phi$ (thus $|b\rangle$ and $|d\rangle$) time independent and use $\Omega(t)$ and $\phi_{1}(t)$ as adjustable parameters for designing evolutions. 

An arbitrary evolution of the system can be formulated as $\left[\left|\psi_{0}(t)\right\rangle,\left|\psi_{1}(t)\right\rangle,\left|\psi_{2}(t)\right\rangle\right]=U(t,0)\left(|d\rangle,|b\rangle,|e\rangle\right)$ with the evolution operator $U(t,0)=\mathcal{T}e^ {-i \int_{0}^{t} H(t^{\prime}) \mathrm{d} t^{\prime}}=\sum^{2}_{m=0}|\psi_{m}(t)\rangle\langle \psi_{m}(0) |$, where $\mathcal{T}$ stands for time ordering. In NHQC schemes~\cite{Sjoqvist2012a, Xu2012Nonadiabatic}, a nonadiabatic cyclic evolution is engineered so that at the end moment $\tau$, $U(\tau,0)\left(|d\rangle,|b\rangle,|e\rangle\right)=\left(|d\rangle,e^{i\gamma}|b\rangle,e^{-i\gamma}|e\rangle\right)$, where $\gamma$ is a geometric phase determined by the path of evolution. When transformed and truncated into the computational subspace, $U(\tau,0)$ has a form of $e^{-i\frac{\gamma}{2} {\bf n} \cdot {\bf {\sigma}}}$, where ${\bf{n}}=(\sin\theta\cos\phi,\sin\theta\sin\phi,\cos\theta)$. Therefore, an arbitrary single-qubit gate can be realized by properly choosing $\theta$, $\phi$, and $\gamma$. 

Conventional NHQC schemes impose the condition of parallel transport, $\left\langle\psi_{m}(t)|H(t)| \psi_{n}(t)\right\rangle=0$ ($m,n=0, 1$), to ensure a geometric phase (i.e., the accumulated dynamical phase is zero) and the robustness of gates. However, it was proved that the infidelity of the resulting NHQC gates showed an identical dependence on control errors to the second order as a dynamical gate~\cite{Zheng2016,Xu2017}. In other words, the NHQC gates do not exhibit a better robustness against control errors as expected. This puzzle has recently been resolved by some authors of this work. Liu \textit{et al.}~\cite{Liu2020} showed that the condition of parallel transport given above cannot alone guarantee that the resulting phases are pure geometric. Specifically, the ``geometric" phases may become contaminated by a residual dynamical phase due to a nonzero cross-coupling of $\left\langle\psi_{1}(t)|H(t)| \psi_{2}(t)\right\rangle$. This fact compromises the prerequisite for the robustness of geometric gates.

A solution of this issue is to impose the condition~\cite{Liu2020}
\begin{equation}\label{Condition}
D_{mn}\equiv\int_{0}^{\tau}d_{mn}(t) d t=0, \quad m, n=0, 1, 2 ,
\end{equation}
where $d_{mn}(t) = \bra{\psi_{m}(t)}H(t) \ket{\psi_{n}(t)}$. For $m,n=0, 1$, this condition represents a relaxed version of the abovementioned condition of parallel transport. On the other hand, $D_{12}=0$ ensures that the dynamical phase due to a nonzero $\left\langle\psi_{1}(t)|H(t)| \psi_{2}(t)\right\rangle$ amounts to zero in a cyclic evolution, resulting in a pure geometric phase. SR-NHQC gates constructed following Eq.(\ref{Condition}) exhibit an enhanced robustness against control errors or similar imperfections compared to conventional NHQC gates and the standard dynamical gates. In Appendix \ref{B5}, we show that for an error of the form $H^{\prime}(t)=(1+\varepsilon) H(t)$ [i.e., an error in the driving amplitude as $\varepsilon\Omega(t)$, the fidelity of the SR-NHQC gate is given by:
\begin{equation}\label{gateF}
F(\varepsilon) = \sqrt{\cos ^{2} \frac{\gamma}{2}+\sin ^{2} \frac{\gamma}{2} \cos ^{4} \frac{\pi \varepsilon}{2}\left(1+\sin ^{2} \frac{\pi \varepsilon}{2}\right)^{2}} \ .
\end{equation}
In the limit of $|\varepsilon| \ll 1$, $F(\varepsilon) \approx 1-\pi^{4}\varepsilon^4(1-\cos\gamma)/32$, exhibiting a  \emph{fourth}-order dependence on the error instead of second-order dependence.

When expressing the Hamiltonian $H(t)$ as a function of $\Omega(t)$ and $\phi_1(t)$, we have  many different choices of these adjustable parameters to fulfill the constraint of Eq.~(\ref{Condition}). Here for simplicity, we use a pulse sequence of six segments to construct the single-qubit SR-NHQC gates with a total gate time of $120$~ns (details are given in Appendix \ref{B1}). Each segment contains a pair of microwave drives on resonance with the $\ket{g} \leftrightarrow \ket{e}$ and $\ket{e} \leftrightarrow \ket{f}$ transitions. The operation of each segment corresponds to a rotation in the subspace of \{$\ket{b}$, $\ket{e}$\} as shown in Fig.~\ref{fig1}(b). In the computational subspace $\{|g\rangle,|f\rangle\}$, the evolution operator $U(\tau,0)$ has the form $U_1(\theta,\phi,\gamma)=e^{-i\frac{\gamma}{2} {\bf n} \cdot {\bf {\sigma}}}$ with ${\bf{n}}=(\sin\theta\cos\phi,\sin\theta\sin\phi,\cos\theta)$, similar to the case of conventional NHQC discussed above.

\section{SINGLE-QUBIT SR-NHQC GATE EXPERIMENT}
The single-qubit SR-NHQC experiment is carried out using a superconducting circuit, where a superconducting transmon qubit~\cite{Koch2007} is dispersively coupled to two microwave cavities~\cite{Paik2011, Vlastakis2013, Kirchmair2013, Sun2014}, one for readout of the qubit state and the other for storage of a microwave photonic qubit. The frequencies of the $\ket{g} \leftrightarrow \ket{e}$ and $\ket{e} \leftrightarrow \ket{f}$ transitions of the transmon qubit are {$\omega_{ge}/2\pi = 5.31$~GHz and $\omega_{ef}/2\pi = 5.12$~GHz}, respectively. The readout cavity with a transition frequency of {$\omega_R/2\pi = 8.68$~GHz} is used to perform high-fidelity and simultaneous readout of the $\ket{g}$, $\ket{e}$, and $\ket{f}$ states. The storage cavity has a transition frequency of {$\omega_S/2\pi = 6.56$~GHz} and allows for the implementation of two-qubit SR NHQC between the transmon states \{$\ket{g}$, $\ket{f}$\} and the Fock states \{$\ket{0}$, $\ket{2}$\} of the photonic qubit. More details of the device and measurement setup can be found in Appendix \ref{A}.

\begin{figure}[t]
	\includegraphics{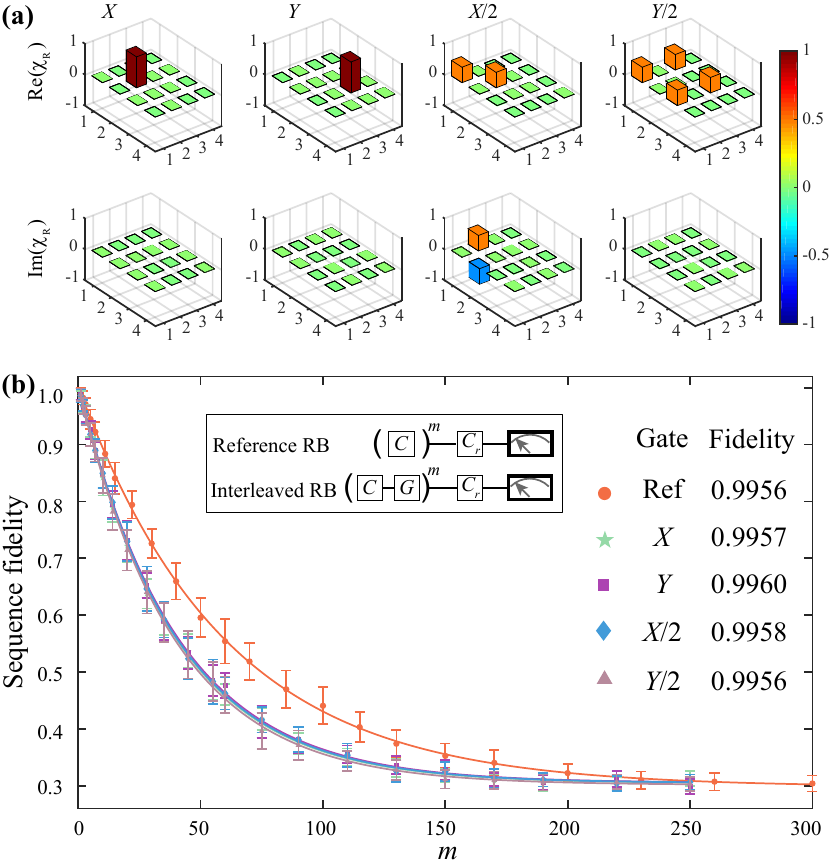}
	\caption{Characterization of the single-qubit SR-NHQC gates by QPT and RB methods.
	(a) Bar charts of the real and imaginary parts of the reduced quantum process matrices $\chi_\mathrm{R}$ of the four specific holonomic gates $X$, $Y$, $X/2$ and $Y/2$, respectively. The numbers along the $x$ and $y$ axes represent operators in the set of \{$I$, $\sigma_x$, $-i\sigma_y$, $\sigma_z$\} acting on the $\{\ket{g},\ket{f}\}$ subspace. The solid black outlines are for ideal gates. The colorbar indicates the magnitude of the real and imaginary parts of the reduced quantum process matrices with dimensionless unit.
	(b) Sequence fidelity as a function of the number of Clifford gates $m$ for both the reference and interleaved RB experiments. Each data point is averaged over 50 random sequences, with the standard deviations plotted as error bars. Fitting the reference curve gives an average gate fidelity of 0.9956 for the single-qubit SR-NHQC gates. Fidelity of the four specific gates $X$, $Y$, $X/2$, and $Y/2$ can be extracted from the difference between the reference and the interleaved decay curves. Inset shows the pulse sequences for the RB experiments.}
\label{fig2}
\end{figure}

We first demonstrate the tunability of $\theta$, $\phi$, and $\gamma$ for realizing arbitrary single-qubit gates. Figures~\ref{fig1}(c)-(e) show the measured population of qubit states as a function of $\theta$, $\phi$, and $\gamma$ for the three representative gate sets $U_1(\theta, 0, \pi/2)$, $U_1(\pi/2,\phi,\pi/2)$, and $U_1(\pi/2,0,\gamma)$, respectively. The experimental results agree well with our numerical simulations. We then characterize the single-qubit SR-NHQC gates using quantum process tomography (QPT) including all three states of \ket{g}, \ket{e}, and \ket{f} (details are given in Appendix \ref{B2}). Figure~\ref{fig2}(a) shows the reduced quantum process matrix in the subspace of \{$\ket{g}$, $\ket{f}$\} for the four specific gates $X = U_1(\pi/2, 0, \pi)$, $Y = U_1(\pi/2,\pi/2,\pi)$, $X/2=U_1(\pi/2, 0, \pi/2)$, and $Y/2=U_1(\pi/2, \pi/2, \pi/2)$. The average process fidelity is 0.9858. We also use the Clifford-based RB~\cite{RBSQProtocol, RBMultiQProtocol, RBInterleaved} to characterize the single-qubit SR-NHQC gates. The reference RB experiment gives an average gate fidelity of 0.9956 for the single-qubit SR-NHQC gates in the Clifford group. The difference between the reference and interleaved RB experiments gives the gate fidelities 0.9957, 0.9960, 0.9958, and 0.9956 for the four SR-NHQC gates of $X$, $Y$, $X/2$, and $Y/2$, respectively. Infidelities of these gates mainly come from the decoherence of both $\ket{e}$ and $\ket{f}$ states of the qubit with a contribution of $4.3\times 10^{-3}$ to gate errors (see Appendix \ref{D}). 

\begin{figure}[t]
	\includegraphics{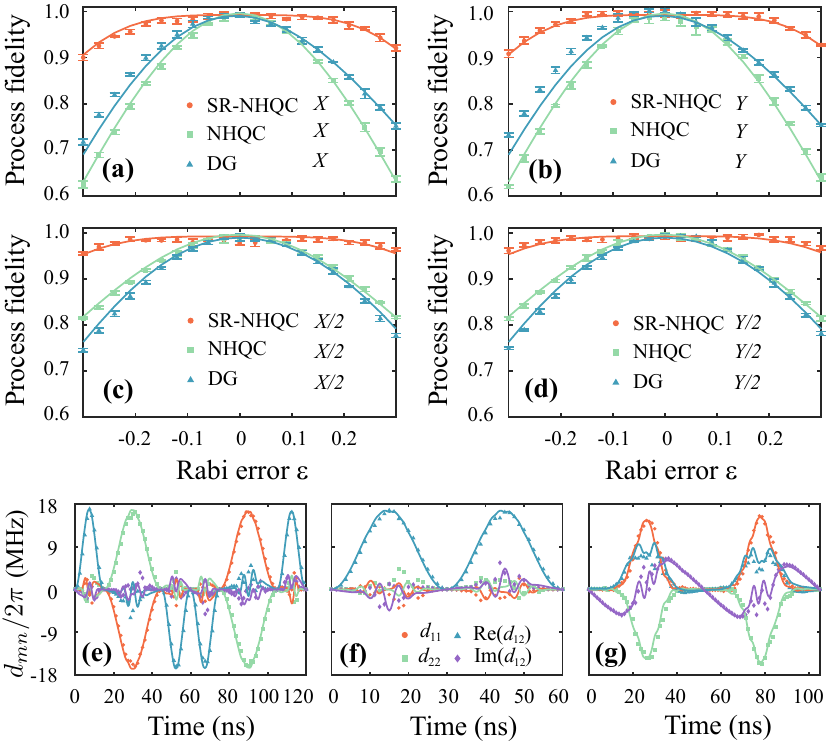}
	\caption{Robustness of the single-qubit SR-NHQC gates. (a)-(d) Process fidelity as a function of the Rabi error for single-qubit gates $X$, $Y$, $X/2$, and $Y/2$ realized by SR NHQC, conventional NHQC, and dynamical means. (e)-(g) Directly measured dynamical phases accumulated over time for an $X$ gate realized by SR NHQC, conventional NHQC, and dynamical means, respectively. Symbols and solid lines are experimental results and numerical simulations.}
	\label{fig3}
\end{figure}

Next, we demonstrate the enhanced robustness of the SR-NHQC gates by comparing them to the conventional NHQC gates and the standard dynamical gates. Using QPT measurements, we study the performance of the four specific gates $X$, $Y$, $X/2$, and $Y/2$ realized via the three different schemes in the presence of a Rabi error $\varepsilon$. The measured process fidelities of these four gates as a function of $\varepsilon$ are shown in Figs.~\ref{fig3}(a)-(d). The SR-NHQC gates are clearly superior to the other two types of gate in terms of robustness. 

As discussed above, the super robustness of the SR-NHQC gates is guaranteed by a suppression of accumulated dynamical phases via imposing Eq.~(\ref{Condition}). In order to examine whether this condition is satisfied, we directly measure the accumulation rates of dynamical phases given by $d_{mn}(t)$ for SR NHQC, conventional NHQC, and dynamical gates (details are given in Appendix \ref{B4}). For the case of an $X$ gate, the results are shown in Figs.~\ref{fig3}(e)-(g). Integrating the measured $d_{mn}(t)$ gives the total dynamical phases $D_{11}$, $D_{22}$, and $D_{12}$, which are $-0.01\pi$, $0.02\pi$, $(-0.01-0.05i)\pi$ for the SR-NHQC $X$ gate, $-0.06\pi$, $0.08\pi$, $(0.97+0.038i)\pi$ for the conventional NHQC $X$ gate, and $0.78\pi$, $-0.78\pi$, $0.47\pi$ for the dynamical $X$ gate, respectively. Thus, we have indeed verified that Eq.~(\ref{Condition}) holds for the SR-NHQC gate. Specifically, the suppression of $D_{12}$ is critical for achieving the super robustness of the SR-NHQC scheme, which is not guaranteed in the conventional NHQC schemes. 

\begin{figure}[t]
	\includegraphics{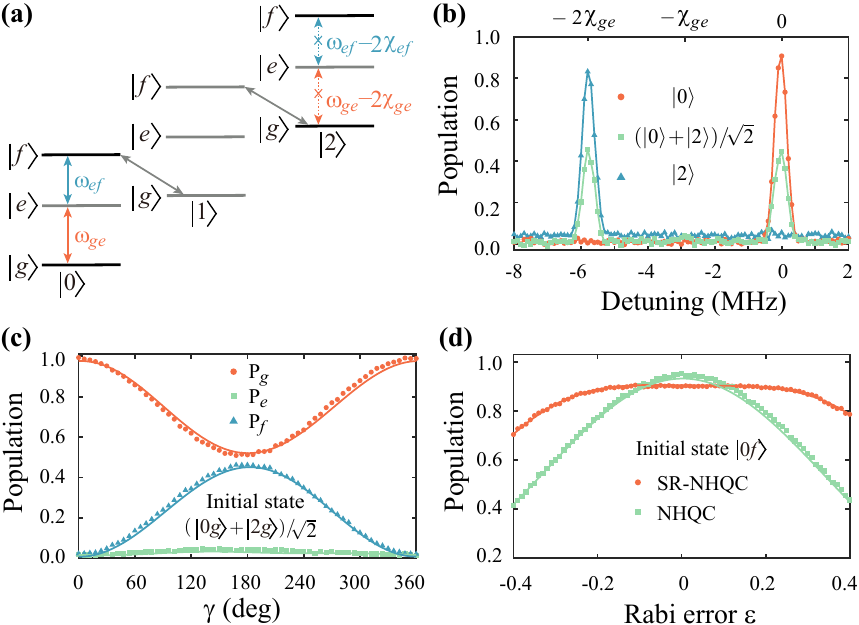}
	\caption{Two-qubit SR-NHQC gates.
	(a) Energy diagram of the coupled transmon-photonic-qubit system for implementing two-qubit SR-NHQC gates. Two microwave drives on resonance with the $\ket{0g}\leftrightarrow \ket{0e}$ and $\ket{0e}\leftrightarrow\ket{0f}$ transitions are used to generate arbitrary rotations in the \{$\ket{0g}$, $\ket{0f}$\} subspace, while keeping $\ket{2g}$ and $\ket{2f}$ unaffected due to the strong dispersive couplings $\chi_{ge}$ and $\chi_{ef}$ between the transmon and photonic qubits. 
	(b) Spectroscopy of the transmon qubit corresponding to the three Fock states $\ket{0}$, $\ket{2}$, and $(|0\rangle+|2\rangle)/\sqrt{2}$ of the photonic qubit. The Fock states are generated by sequentially applying Raman drives on the transitions of $\ket{0f} \leftrightarrow \ket{1g}$ and $\ket{1f} \leftrightarrow \ket{2g}$ (solid gray arrow lines in (a)). 
	(c) Populations of the transmon qubit as a function of $\gamma$ for the two-qubit SR-NHQC gate $U_2(\pi/2,0,\gamma)$ with an initial state $\left(\ket{0g}+\ket{2g}\right)/\sqrt{2}$. Symbols and lines are experimental data and numerical simulations, respectively. 
	(d) Population of the ground state of the transmon qubit as a function of the Rabi error for a two-qubit controlled-NOT (CNOT) gate realized by SR-NHQC and conventional NHQC means. The initial state is $\ket{0f}$ in both cases.}
	\label{fig4}
\end{figure}

\section{TWO-QUBIT SR-NHQC GATE}
We also demonstrate nontrivial two-qubit SR-NHQC gates between a superconducting transmon qubit and a photonic qubit, using the scheme illustrated in Fig.~\ref{fig4}(a). In our experiment, two microwave drives are applied on resonance with the $\ket{0g}\leftrightarrow \ket{0e}$ and $\ket{0e}\leftrightarrow\ket{0f}$ transitions. Here the numbers represent the Fock states of the photonic qubit. Similar to the single-qubit case, we thus realize an arbitrary holonomic gate $U_1(\theta, \phi, \gamma)$ with a total gate time of $2.76~\mu$s in the \{$\ket{0g}$, $\ket{0f}$\} subspace, while keeping $\ket{2g}$ and $\ket{2f}$ unaffected thanks to the strong dispersive $ZZ$ interaction (about 2~MHz). This operation thus creates a control gate in the two-qubit subspace of \{$\ket{0g}$, $\ket{0f}$, $\ket{2g}$, $\ket{2f}$\} described by
\begin{equation}
\label{U2}
U_2(\theta,\phi,\gamma) = \left(
\begin{array}{cc}
   U_1(\theta,\phi,\gamma)   &   0   \\
   0   &   I
\end{array}
\right).
\end{equation}

\noindent Of course, the strong $ZZ$ interaction between the two qubits also induces a conditional phase operation, which renders the overall evolution matrix to be different from Eq.~(\ref{U2}). However, since this interaction is time independent, the resulting phase can always be zeroed by properly setting the overall evolution time. Therefore, we can safely neglect this phase in the following discussion.  

In order to characterize the two-qubit SR-NHQC gates, we first prepare different initial Fock states of the photonic qubit by sequentially applying two Raman transition drives~\cite{Pechal2014, Zeytinoglu2015} for $\ket{0f} \leftrightarrow \ket{1g}$ and $\ket{1f} \leftrightarrow \ket{2g}$, respectively, with the experimental results shown in Fig.~\ref{fig4}(b). Figure ~\ref{fig4}(c) plots the populations of the transmon qubit as a function of $\gamma$ for the two-qubit SR-NHQC gate set $U_2(\pi/2,0,\gamma)$ applied to the initial state of $(\ket{0g}+\ket{2g})/\sqrt{2}$. A gate fidelity of 0.944 is estimated for the realized two-qubit CNOT gate defined by $U_2(\pi/2,0, \pi)$ (details are given in Appendix \ref{C2}). The infidelity of this two-qubit SR-NHQC gate mainly comes from the decoherence of the qubits due to a long gate operation time  with a coherence-limited error of $5.8\times 10^{-2}$ (see Appendix \ref{D}). A higher fidelity can be achieved with a shorter gate operation time (about 60 ns) by utilizing a tunable coupler between two superconducting transmon qubits, where a typical $ZZ$ coupling between qubits of the order of 100~MHz can be routinely achieved~\cite{Xu2020, Collodo2020}. We further demonstrate the super robustness of the two-qubit SR-NHQC CNOT gate by measuring the populations as a function of the Rabi error $\varepsilon$ and comparing to the conventional NHQC two-qubit CNOT gate, as shown in Fig.~\ref{fig4}(d). The experimental results clearly show the superior robustness of the SR-NHQC CNOT gate. Again, the under performance of the SR-NHQC gate at small $\varepsilon$ is due to a longer gate operation time and can be greatly improved when implemented on a system of two transmon qubits and a tunable coupler.

\section{CONCLUSION}
In summary, we have experimentally demonstrated a universal gate set based on a super robust NHQC scheme in an architecture of circuit quantum electrodynamics. Compared to conventional NHQC schemes, the SR-NHQC scheme guarantees an enhanced robustness against quasistatic errors appearing in the transverse direction (e.g., Rabi error in the $xy$ control pulses) by imposing additional constraints that help suppress dynamical phases. The realized single- and two-qubit SR-NHQC gates achieve an average fidelities of 0.9956 and 0.944, respectively, and both show super robustness against Rabi errors, as predicted. In addition, for the single-qubit SR-NHQC gates, we have directly measured the residual dynamical phases and verified the suppression of such phases inherent in our scheme. Given its generality and simplicity, the SR-NHQC scheme can be implemented on other platforms such as trapped ions, quantum dots, Rydberg atoms, and nuclear magnetic resonance, etc. Our work thus paves a way to construct universal super robust holonomic quantum gates for future large-scale quantum computation. 

\begin{acknowledgments}
This work is supported by the Key-Area Research and Development Program of Guangdong Province (Grant No. 2018B030326001), the National Natural Science Foundation of China (Grants No. 11904158 and No. 11874156), the Guangdong Provincial Key Laboratory (Grant No. 2019B121203002), the Guangdong Innovative and Entrepreneurial Research Team Program (Grant No. 2016ZT06D348), the Natural Science Foundation of Guangdong Province (Grant No. 2017B030308003), the Science, Technology and Innovation Commission of Shenzhen Municipality (Grants No. JCYJ20170412152620376 and No. KYTDPT20181011104202253), and Shenzhen-Hong Kong cooperation zone for technology and innovation (Contract No. HZQB-KCZYB-2020050).
\end{acknowledgments}

\appendix

\section{EXPERIMENTAL DETAILS}\label{A}
\subsection{Device and setup} \label{A1}

\begin{figure}[t]
	\centering
	\includegraphics{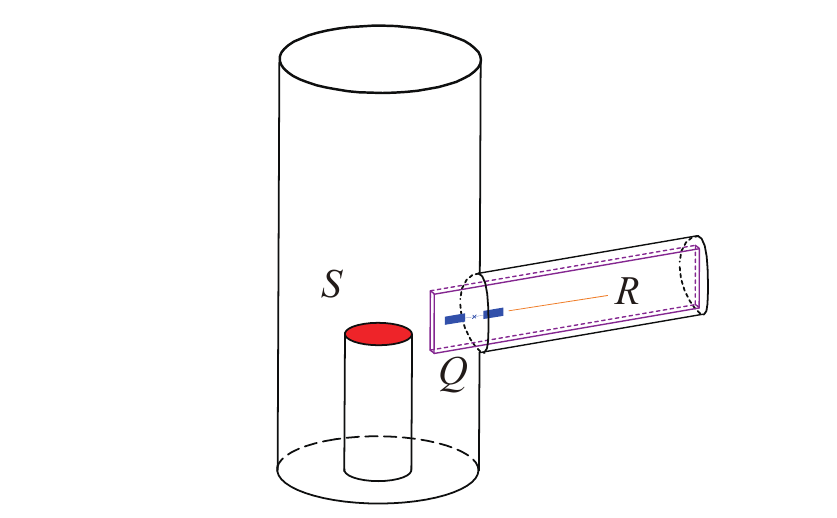}
	\caption{Schematic diagram of the device. A superconducting transmon qubit~($Q$) is dispersively coupled to two 3D microwave cavities. One is used for readout~($R$) of the qubit states, and the other for storage~($S$) of a photonic qubit and implementing the two-qubit SR-NHQC gates.}
	\label{Figs1}
\end{figure}

\begin{figure}[b]
	\centering
	\includegraphics{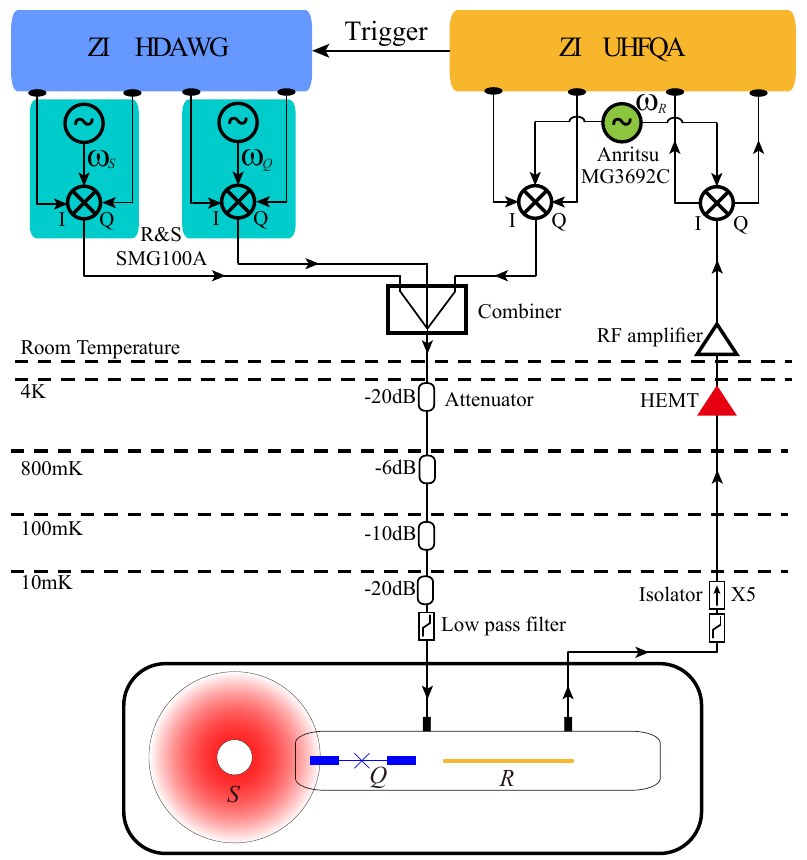}
	\caption{The full wiring of the experimental setup.}
	\label{Figs2}
\end{figure}

The experimental device with a circuit quantum electrodynamics~(cQED) architecture~\cite{wallraff2004strong, blais2004cqed} contains a three-dimensional (3D) coaxial stub cavity as the storage cavity~($S$), a superconducting transmon qubit~($Q$), and a stripline readout resonator~($R$). The schematic diagram of the device is shown in Fig.~\ref{Figs1}. The 3D coaxial stub cavity is first machined from a single block of high purity aluminum and then chemically etched in order to improve the quality factor of the microwave cavity~\cite{Reagor2013}. The superconducting transmon qubit containing a Josephson junction connected to two antenna pads is fabricated on a $c$-plane sapphire chip with a double angle evaporation of aluminum after a single electron-beam lithography step. The qubit chip is placed inside a horizontal tunnel of the 3D device, with the two antenna pads coupling to the coaxial stub cavity mode and the stripline readout resonator mode, respectively~\cite{Axline2016, Reagor2016}. The stripline resonator is formed by the metal wall of the tunnel and an aluminum stripline, which is simultaneously patterned on the same chip with the superconducting qubit.

The experimental device is placed inside a magnetic shield and mounted in a dilution refrigerator with a base temperature of about 10~mK. The full wiring of the experimental setup is shown in Fig.~\ref{Figs2}. Attenuators and filters are used on the microwave lines to reduce the radiation noises of the signals. Both the qubit and storage cavity control pulses are generated by IQ modulations with two analogy channels of a Zurich Instruments high-density arbitrary waveform generator (ZI HDAWG) and a mixer. All the qubit control pulses for the initial state preparations and prerotations before measurement have a cosine-shape envelope with a duration of 30~ns. The technique of "derivative removal by adiabatic gate" (DRAG) is applied to both $|g\rangle \leftrightarrow |e\rangle$ and $|e\rangle \leftrightarrow |f\rangle$ transitions in order to suppress the leakage to undesired energy levels~\cite{Motzoi2009, Gambetta2011}. The readout pulse is generated by in-phase and quadrature (IQ) modulations of two analog channels of a Zurich Instruments ultra-high-frequency quantum analyzer (ZI UHFQA) with a signal generator as the local oscillator. The readout signal is amplified by a high electron mobility transistor~(HEMT) at the 4K stage and a standard commercial amplifier at room temperature, and does a down-conversion with the same local oscillator as generating the readout pulse. Finally, the readout signal is digitized and recorded by the same ZI UHFQA.

\begin{table}[tb]
\caption{Hamiltonian parameters.}
\begin{tabular}{p{2cm}<{\centering}p{2cm}<{\centering}p{2cm}<{\centering}p{2cm}<{\centering}}  
  \hline
  \multicolumn{2}{c}{Frequencies (GHz)} & \multicolumn{2}{c}{Couplings (MHz)} \\
  \hline
  &&&\\[-0.9em]
  $\omega_{R}/2\pi$ & 8.68  & $\chi_{{ge}}^{{RQ}}/2\pi$ & 2.52  \\
    &&&\\[-0.9em]
  $\omega_{S}/2\pi$ & 6.56  & $\chi_{{ef}}^{{RQ}}/2\pi$ & 2.39   \\
  &&&\\[-0.9em]
  $\omega_{{ge}}/2\pi$ & 5.31  & $\chi_{{ge}}^{{SQ}}/2\pi$ & 2.87  \\
  &&&\\[-0.9em]
  $\omega_{{ef}}/2\pi$ & 5.12  & $\chi_{{ef}}^{{SQ}}/2\pi$ & 2.08  \\
  &&&\\[-0.9em]
  \hline
\end{tabular}\vspace{-6pt}
\label{TableS1}
\end{table}

\subsection{System Hamiltonian and coherence properties}\label{A2}
In our device, a superconducting transmon qubit is dispersively coupled to two cavity modes: a storage cavity mode and a readout cavity mode. The transmon qubit has a large anharmonicity and is considered as a three-level artificial atom, while each cavity mode is considered as a harmonic oscillator. Thus, the Hamiltonian of the whole system can be described as
\begin{eqnarray} 
\label{Hamiltion}
\mathcal{H} &=& \omega_R\left(a^\dagger_R a_R +1/2\right)+\omega_S\left(a^\dagger_S a_S +1/2\right) \notag \\ 
 &+& \omega_{ge}|e\rangle\langle e|+ \left(\omega_{ge}+\omega_{ef}\right)|f\rangle\langle f| \notag \\
&-& \chi^{RQ}_{ge}|e\rangle\langle e| a^\dagger_R a_R - \left(\chi^{RQ}_{ge}+\chi^{RQ}_{ef}\right)|f\rangle\langle f| a^\dagger_R a_R \notag \\
&-& \chi^{SQ}_{ge}|e\rangle\langle e| a^\dagger_S a_S - \left(\chi^{SQ}_{ge}+\chi^{SQ}_{ef}\right)|f\rangle\langle f| a^\dagger_S a_S, \notag \\
\end{eqnarray}
where $\omega_{R,S}$ are the resonant frequencies of the readout and the storage cavities, respectively; $a_{R,S}$~($a^{\dagger}_{R,S}$) are the corresponding ladder operators; $\omega_{ge}$ and $\omega_{ef}$ are transition frequencies among the lowest three energy levels $\{|g\rangle,|e\rangle,|f\rangle \}$; and the $\chi$ are corresponding dispersive couplings between the qubit and the cavity modes.

\begin{figure}[t]
	\centering
	\includegraphics{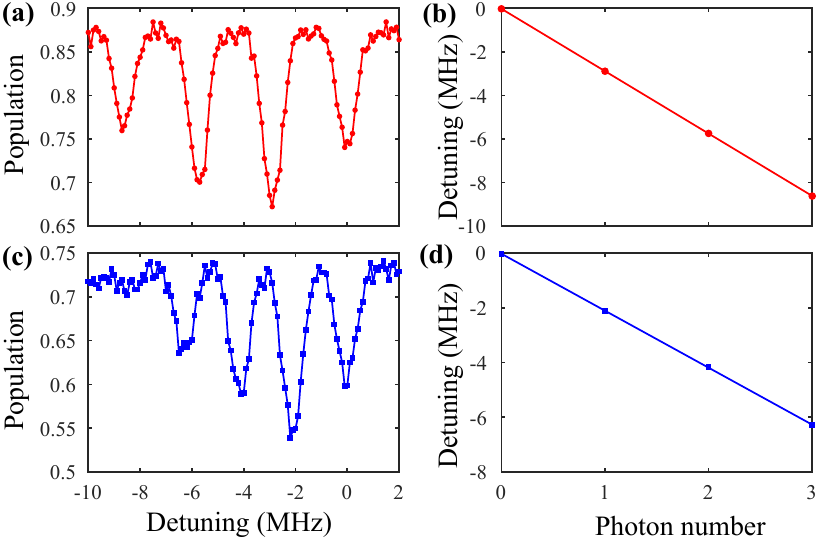}
	\caption{Calibration of dispersive couplings between the qubit and the storage cavity mode with number splitting experiments. (a) Qubit $\ket{g} \leftrightarrow \ket{e}$ transition frequency spectroscopy with a coherent state in the storage cavity mode. (b) The resonant qubit $\ket{g} \leftrightarrow \ket{e}$ transition frequencies as a function of the photon numbers in the storage cavity. Solid line is a linear fit. (c) Qubit $\ket{e} \leftrightarrow \ket{f}$ transition frequency spectroscopy with a coherent state in the storage cavity mode. (d) The resonant qubit $\ket{e} \leftrightarrow \ket{f}$ transition frequencies as a function of the photon numbers in the storage cavity. Solid line is a linear fit.}
	\label{Figs3}
\end{figure}

\begin{table}[b]
\caption{Coherence properties of the system.}
\begin{tabular}{p{2.6cm}<{\centering} p{2.6cm}<{\centering} p{2.6cm}<{\centering} }
  \hline
  Modes & $T_1$ & $T_2^*$ \\\hline
  &&\\[-0.9em]
  Readout cavity & 66~ns & -\\
  &&\\[-0.9em]
  Storage cavity & 334~$\mu$s & 243~$\mu$s\\
  &&\\[-0.9em]
  Qubit $\ket g \leftrightarrow \ket e$ & 18.9~$\mu$s & 25.9~$\mu$s  \\
  &&\\[-0.9em]
  Qubit $\ket e \leftrightarrow \ket f$ & 12.7~$\mu$s & 12.9~$\mu$s  \\
  &&\\[-0.9em]
  \hline
\end{tabular}
\label{TableS2}
\end{table}

All the parameters in the previous Hamiltonian are calibrated with the standard cQED technique. The dispersive couplings $\chi_{ge}^{SQ}$ and $\chi_{ef}^{SQ}$ between the qubit and the storage cavity mode are calibrated with a number splitting experiment through qubit $\ket{g} \leftrightarrow \ket{e}$ and $\ket{e} \leftrightarrow \ket{f}$ transition frequency spectroscopes with a coherent state in the storage cavity \cite{Schuster2007photon}. The experiment results are shown in Fig.~\ref{Figs3}. The dispersive couplings $\chi_{ge}^{RQ}$ and $\chi_{ef}^{RQ}$ between the qubit and the readout cavity are calibrated through a readout frequency spectroscopy experiment with the qubit prepared in different initial states $\ket{g}$, $\ket{e}$, and $\ket{f}$, respectively. The experimental results are shown in Fig.~\ref{Figs5}(a). All the parameters in Eq.~\ref{Hamiltion} are measured and listed in Table.~\ref{TableS1}. 

\begin{figure}[t]
	\centering
	\includegraphics{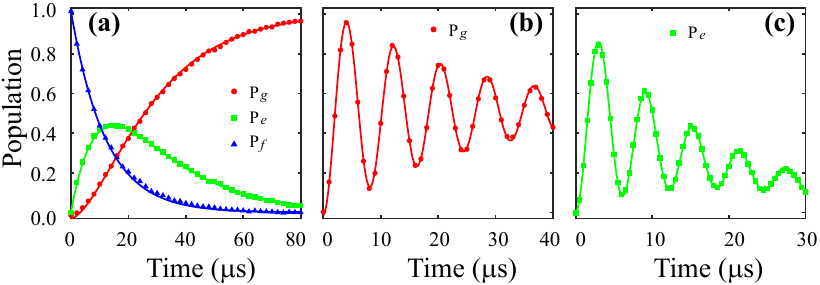}
	\caption{Coherence properties of the three-level transmon qubit. (a) Population decay curves with the transmon qubit initialized in the $|f\rangle$ state with a global fitting based on the rate equation (solid lines). (b-c) Ramsey oscillation experiment between $\ket{g}$ and $\ket{e}$ states~(b), and $\ket{e}$ and $\ket{f}$ states~(c) of the transmon qubit, respectively. Solid lines are corresponding fittings with an exponentially damped sinusoidal function. Note that the offset of the fitting function for the $\ket{e} \leftrightarrow \ket{f}$ Ramsey experiment is an exponential decay term to account for both $\ket{e}$ and $\ket{f}$ states decaying to the $\ket{g}$ state.}
	\label{Figs4}
\end{figure}

\begin{figure}[t]
	\centering
	\includegraphics{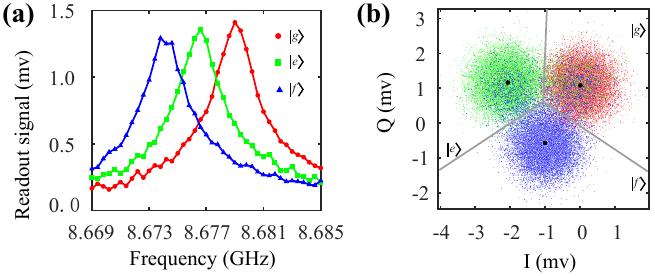}
	\caption{Readout characterization of the $\ket{g}$, $\ket{e}$, and $\ket{f}$ states of the transmon qubit. (a) Readout frequency spectroscopy for the qubit initially prepared in the $|g\rangle$, $|e\rangle$ and $|f\rangle$ states, respectively. (b) Recorded data points in the I-Q plane in a single shot experiment for qubit initially prepared in $|g\rangle$, $|e\rangle$ and $|f\rangle$ state, respectively. The I-Q plane is divided into three regions with the black dot in each region corresponding to the median of that distribution. The three regions are separated by three solid gray lines to distinguish the qubit states and obtain the assignment probabilities.}
	\label{Figs5}
\end{figure}

In addition, the coherence properties of the whole system are also experimentally measured and listed in Table~\ref{TableS2}. The relaxation time of the readout cavity is extracted from the linewidth of the readout frequency spectroscopy. The relaxation times of the $\ket{e}$ and $\ket{f}$ states of the transmon qubit are obtained by measuring the free evolutions of the populations $P_g$, $P_e$, and $P_f$ with an initial $|f\rangle$ state, following the technique described in Ref.\cite{Peterer2015}. The populations $P_e$ and $P_f$ are measured by mapping them onto the population of ground state $|g\rangle$ through a $\pi$ pulse and two sequential $\pi$ pulses, respectively. The experimental results are shown in Fig.~\ref{Figs4}(a). The decay curves are globally fitted with the rate equation $ d \vec{p}/dt = \Gamma\cdot \vec{p}$, where $\vec{p}= (P_g, P_e, P_f)^T$, and the decay rate matrix $\Gamma$ is
\begin{eqnarray}
\Gamma=\left(
\begin{array}{ccc}
0  & \Gamma_{ge} & \Gamma_{gf} \\
0  & -\Gamma_{ge} & \Gamma_{ef}\\
0  & 0 & -(\Gamma_{gf}+\Gamma_{ef})
\end{array}
\right)\label{decayE}
\end{eqnarray}
where the negligible upward transition rates are ignored and only the downward transition rates $\Gamma_{ge}$, $\Gamma_{ef}$, and $\Gamma_{gf}$ are considered. We note that the transition $|f\rangle\rightarrow|g\rangle$ is forbidden from parity considerations for a single-junction transmon qubit. However, in a realistic device, there are small nonsequential decay rates, which are mainly dominated by some nonquasiparticle processes, such as dielectric loss or coupling to other cavity modes~\cite{Peterer2015}. Since the nonsequential decay rate $\Gamma_{gf}$ is much slower than the sequential decay rates $\Gamma_{ge}$ and $\Gamma_{ef}$, the corresponding relaxation times $1/\Gamma_{ge}$ and $1/\Gamma_{ef}$ of qubit are listed as $T_1$ in Table~\ref{TableS2}.

Then, we measure the dephasing rates between $|g\rangle$ and $|e\rangle$ states, and between $|e\rangle$ and $|f\rangle$ states of the qubit with Ramsey interference experiments; the results are shown in Figs.~\ref{Figs4}(b,c). The Ramsey fringes are fitted with an exponentially damped sinusoidal function $y = y_0+e^{-t/T_2^*}A\cos{(2\pi f t+\varphi)}$ and the extracted $T_2^*$ are also listed in Table~\ref{TableS2}. 

The coherence times $T_1$ and $T_2^*$ of the storage cavity are measured through the relaxation of Fock state $|1\rangle$ and the dephasing of $(|0\rangle+|1\rangle)/\sqrt{2}$, respectively~\cite{Reagor2016}. Both initial states are generated with selective number-dependent arbitrary phase gates~\cite{Heeres2015}.

\subsection{Qubit readout}\label{A3}
Here, we probe the qubit states $\ket{g}$, $\ket{e}$, and $\ket{f}$ via the dispersive readout technique~\cite{blais2004cqed}. We first perform the readout cavity frequency spectroscopy experiment with different qubit initial states $\ket{g}$, $\ket{e}$, and $\ket{f}$. The experimental results are shown in Fig.~\ref{Figs5}(a). We optimize the readout pulse amplitude, duration, frequency, record pulse delay, and integration length to maximize the readout discrimination of all the $\ket{g}$, $\ket{e}$, and $\ket{f}$ states simultaneously. We perform the single-shot experiments with 20 000 repetitions for each initial qubit state, and record the I and Q quadratures of the readout signals with the experimental results shown in Fig.~\ref{Figs5}(b). The I-Q plane is divided into three regions corresponding to the assignments of $\ket{g}$, $\ket{e}$, and $\ket{f}$ states. By counting the number of I-Q data points in the three regions, we could obtain the assignment probability $\vec{P} = (P_0, P_1, P_2)^T$ corresponding to that initial basis state. After repeating the experiments for the three initial qubit states, we could obtain an assignment probability matrix $\mathcal{M}$ in span $\{|g\rangle,|e\rangle, |f\rangle\}$ with
\begin{eqnarray}
\mathcal{M}=\left(
\begin{array}{ccc}
0.942  & 0.080 & 0.076 \\
0.040  & 0.908 & 0.077\\
0.018  & 0.012 & 0.847
\end{array}
\right),\label{Mmatrix}
\end{eqnarray}
where each column represents the qubit assignment probabilities after preparing the qubit in the corresponding basis state. Then the readout errors can be corrected by multiplying the inverse of the assignment matrix $\mathcal{M}$ with the measured probability $\vec{P}$. Therefore, $\vec{P}_\mathrm{corr} = \mathcal{M}^{-1}\cdot \vec{P}$ represents the actual occupation probabilities of the $\ket{g}$, $\ket{e}$, and $\ket{f}$ states of the transmon qubit. Notably, due to the parameter fluctuations in experimental device, this assignment matrix may be a little different from that in Eq.~\ref{Mmatrix}, thus making the final probabilities slightly over or inadequately corrected.

 \begin{figure}[tb]
	\centering
	\includegraphics{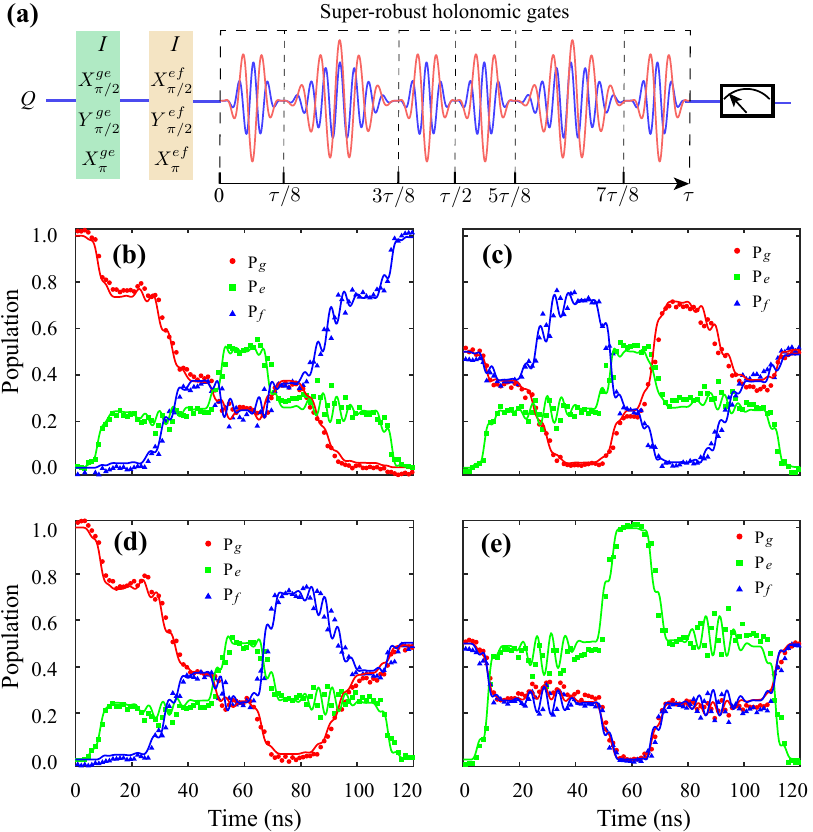}\caption{The experimental pulse sequence (a) and state population evolution of the single-qubit SR-NHQC gates $X$~(b), (c) and $Y/2$~(d), (e) as a function of gate duration for initial states $|g\rangle$~(a,c) and $(|g\rangle-i|f\rangle)/\sqrt{2}$~(b), (d), respectively. Solid lines are corresponding numerical simulations.}
	\label{Figs6}
\end{figure}


\section{SINGLE-QUBIT SR-NHQC GATES}\label{B}
\subsection{State population evolution of single-qubit gates}\label{B1}
Single-qubit SR-NHQC gates are constructed by a pulse sequence of six segments. Each segment contains a pair of microwave drives on resonance with the $\ket{g} \leftrightarrow \ket{e}$ and $\ket{e} \leftrightarrow \ket{f}$ transitions, thus corresponding to a Hamiltonian of $H(t)=\frac{1}{2}\left[\Omega(t)e^{i\phi_{1}(t)}| b\rangle \langle e| +\mathrm{H.c.}\right]$. The operation of each segment corresponds to a rotation $R_\varphi (\vartheta)$ in the equatorial plane of \{$\ket{b}$, $\ket{e}$\} subspace, with rotation angle $\vartheta = \int_0^t{\Omega(t)dt}$ and phase $\varphi = \phi_1$. The total evolution time $(\tau = 120$~ns) is divided into six intervals, with the rotation angle and phase in each segment satisfying
\begin{equation}
    \label{splitpulses}
    \begin{cases}
    \vartheta_1 = \pi/2, \quad \varphi_1 = \gamma-\pi, \quad\quad &t\in[0,\tau/8],\\
    \vartheta_2 = \pi, \quad\quad \varphi_2 = \gamma-\pi/2, \quad &t\in[\tau/8,3\tau/8],\\
    \vartheta_3 = \pi/2, \quad \varphi_3 = \gamma-\pi, \quad\quad &t\in[3\tau/8,\tau/2],\\
    \vartheta_4 = \pi/2, \quad \varphi_4 = 0, \quad\quad\quad\quad &t\in[\tau/2,5\tau/8],\\
    \vartheta_5 = \pi, \quad\quad \varphi_5 = \pi/2, \quad\quad\quad &t\in[5\tau/8,7\tau/8],\\
    \vartheta_6 = \pi/2, \quad \varphi_6 = 0, \quad\quad\quad\quad &t\in[7\tau/8,\tau].\\
    \end{cases}
\end{equation}

In the computational subspace $\{|g\rangle,|f\rangle\}$, the evolution operator $U(\tau,0)$ has form $U_1(\theta,\phi,\gamma)=e^{-i\frac{\gamma}{2} {\bf n} \cdot {\bf {\sigma}}}$, corresponding to a rotation operation around the axis ${\bf{n}}=(\sin\theta\cos\phi,\sin\theta\sin\phi,\cos\theta)$ by an angle of $\gamma$. The total evolution of these six segments satisfies condition~(\ref{Condition}) in the main text, thus corresponding to super robust nonadiabatic holonomic gates with enhanced robustness. We note that this construction of holonomic gates is similar to composite pulses~\cite{Ota2009, Ichikawa2012}, whose robustness also originates from satisfying condition ~(\ref{Condition}) in the main text.

\begin{figure}[t]
	\centering
	\includegraphics{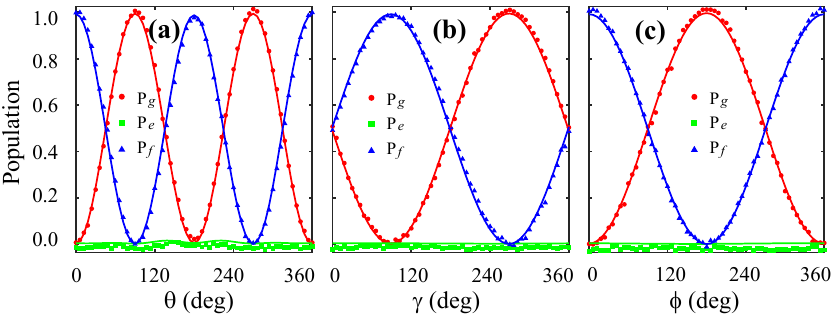}
	\caption{Final state populations of the single-qubit SR-NHQC gates  $U_1(\theta,0,\pi)$ as a function of $\theta$ for initial state $|f\rangle$~(a), $U_1(\pi/2,0,\gamma)$ as a function of $\gamma$ for initial state $(|g\rangle-i|f\rangle)/\sqrt{2}$~(b), and $U_1(\pi/2,\phi,\pi/2)$ as a function of $\phi$ for initial state $(|g\rangle-i|f\rangle)/\sqrt{2}$~(c). Solid lines are corresponding numerical simulations.}
	\label{Figs7}
\end{figure}

In our experiment, each resonant microwave drive of the six segments is implemented with the cosine-shape envelope pulse shown in Fig.~\ref{Figs6}(a), with the pulse amplitudes and phases satisfying Eq.~(\ref{splitpulses}) to achieve better robustness. We first measure the state dynamics of the single-qubit SR-NHQC gates. Firstly, we initialize the qubit with $|g\rangle$ and $(|g\rangle-i|f\rangle)/\sqrt{2}$ states, respectively. Then, six pairs of resonant pulses are applied on the qubit to realize single-qubit SR-NHQC gates $X$ and $Y/2$, respectively. We measure the qubit state populations as a function of the gate duration, with the experimental results given in Figs.~\ref{Figs6}(b-e), which agree well with our numerical simulations. In addition, we demonstrate the arbitrary tunability of the parameters $\theta$, $\phi$, and $\gamma$ for the single-qubit SR-NHQC gates $U_1(\theta, \phi, \gamma)$. Here we present extended data to that in Fig.~1(a) in the main text. Figs.~\ref{Figs7}(a-c) are the measured qubit state populations as a function of $\theta$, $\gamma$, and $\phi$ for the realized single-qubit SR-NHQC gates $U_1(\theta, 0, \pi)$, $U_1(\pi/2,0,\gamma)$, and $U_1(\pi/2,\phi,\pi/2)$, respectively. The experimental results agree well with our numerical simulations, indicating arbitrary control of the super robust single-qubit holonomic gates.

\subsection{Quantum process tomography}\label{B2}
The single-qubit SR-NHQC gates are first characterized by a full quantum process tomography with all the three-level system. We first initialize the three-level transmon qubit with the nine states $\{|g\rangle,|e\rangle,|f\rangle,(|g\rangle+|e\rangle)/\sqrt{2},(|e\rangle+|f\rangle)/\sqrt{2},(|g\rangle+|f\rangle)/\sqrt{2},(|g\rangle-i|e\rangle)/\sqrt{2},(|e\rangle-i|f\rangle)/\sqrt{2},(|g\rangle-i|f\rangle)/\sqrt{2}\}$, then apply the SR-NHQC gates, and finally perform state tomography measurements of the final states. The state tomography measurement requires nine prerotations to reconstruct the density matrix of the three-level qubit state: \{$I$, $X^{ge}_{\pi/2}$, $Y^{ge}_{\pi/2}$, $X^{ge}_{\pi}$, $X^{ge}_{\pi/2}X^{ef}_{\pi}$, $Y^{ge}_{\pi/2}X^{ef}_{\pi}$, $X^{ge}_{\pi}X^{ef}_{\pi/2}$, $X^{ge}_{\pi}Y^{ef}_{\pi/2}$, $X^{ge}_{\pi}X^{ef}_{\pi}$ \}~\cite{ Bianchetti2010}, with the rotation operators read from right to left. The measurements give the result $\langle M_k \rangle = \textrm{Tr}(\rho U^\dagger_kM_IU_k)$ for each prerotation $U_k$ with $k = 0,1,2,...,8$, where $M_I = |g\rangle \langle g| $. The density matrix of the three-level qubit state can then be reconstructed by the maximum likelihood estimation method~\cite{James2001}. With the nine initial states $\rho_i$, the experimental process matrix $\chi_{exp}$ can be extracted from the nine corresponding final states $\rho_f$ through $\rho_f = \sum_{m,n}\chi_{mn}E_m\rho_i E^\dagger_n$~\cite{Nielsen}, where the full set of nine orthogonal basis operators is chosen as \{$I_{gf}$, $\sigma^x_{gf}$, $-i\sigma^y_{gf}$, $\sigma^z_{gf}$, $\sigma^x_{ge}$, $-i\sigma^y_{ge}$, $\sigma^x_{ef}$, $-i\sigma^y_{ef}$, $I_e$ \}~\cite{Abdumalikov2013}, where the $\sigma_{mn}$ are the Pauli operators acting on the $m$ and $n$ energy levels, $I_{gf} = |g\rangle \langle g|+|f\rangle \langle f|$, and $I_e = |e\rangle \langle e|$.

For the single-qubit SR-NHQC gates on the transmon qubit, the state $|e\rangle$ serves as an auxiliary state. Therefore, we have calculated the reduced process matrix $\chi_R$ that describes the process only involving $|g\rangle$ and $|f\rangle$, and ignore any operators acting on the auxiliary state. In order to compare with the process acting on a two-level system, the reduced process matrix $\chi_R$
is obtained by a normalization factor of 3/2. The basis operators of the reduced process matrix are $\{I_{gf}, \sigma^x_{gf},-i\sigma^y_{gf}, \sigma^z_{gf}\}$, or simply $\{I,\sigma_x, -i\sigma_y,\sigma_z\}$ as in the main text. The quantum process fidelity of the corresponding gate is defined as $F = |\mathrm{Tr}(\chi_R \chi_\mathrm{ideal}^\dagger )|$, where $\chi_\mathrm{ideal}$ is the ideal process matrix for the corresponding gate.

\subsection{Randomized benchmarking}\label{B3}
In the single-qubit RB experiment, we perform both the reference RB and interleaved RB experiments with the experimental sequences shown in the inset of Fig. 2(b) in the main text. In the reference RB experiment, we first apply a random sequence of $m$ quantum gates chosen from the single-qubit Clifford group, then append a recovery gate ($C_r$) to invert the whole sequence, and finally measure the ground-state probability as the sequence fidelity. The whole experiment is repeated for $k=50$ different sequences to get the average sequence fidelity. In the interleaved RB experiment, a specific gate $G$ is interleaved into the $m$ random Clifford gates, and a similar recovery gate is applied to invert the whole sequence. The experimentally measured sequence fidelity decay curves as a function of the number of Clifford gates $m$ for both the reference RB and interleaved RB experiments are fitted to $F = A p^m+B$ with different sequence decays $p = p_{\textrm{ref}}$ and $p = p_{\textrm{gate}}$. The average gate fidelity is given by $F_\textrm{ref} = 1-(1-p_{\textrm{ref}})(d-1)/d/1.875$ with $d = 2^N$ for $N$ qubits. Here the number 1.875 accounts for a total 45 physical gates to construct the 24 Clifford gates in the single-qubit Clifford group~\cite{Barends2014Superconducting}. The difference between the reference and interleaved RB experiments gives the specific gate fidelity $F_\textrm{gate} = 1-(1-p_{\textrm{gate}}/p_{\textrm{ref}})(d-1)/d$.

\subsection{Measurement of dynamical phases}\label{B4}
In the main text, we have explicitly measured the accumulated dynamical phase for our SR-NHQC gates, as well as conventional NHQC gates and dynamical gates. In Sec.B 1 above, we introduced the construction of SR-NHQC gates in our scheme with a total evolution time $\tau = 120$~ns. Conventional NHQC gates are easily realized according to Refs.~\cite{Herterich2016,Hong2018} with a total evolution time $\tau = 60$~ns. Dynamical gates are constructed following Ref.~\cite{Liu2019}. A quantum system driven by the Hamiltonian $H(t)=\frac{1}{2}\left[\Omega(t)e^{i\phi_{1}(t)}| b\rangle \langle e| +\mathrm{h.c.}\right]$ can evolve along state $|\psi_{1}(t) \rangle = e^{-if/2}[\cos(\chi/2)e^{-i\varphi/2}|b\rangle+\sin(\chi/2)e^{i\varphi/2}|e\rangle]$, where $f$, $\varphi$ and $\chi$ are time-dependent auxiliary parameters, which satisfy the equations
 \begin{equation}\label{relationships}
\begin{split}
\dot{\varphi}&=-\dot{f}\cos\chi, \\
 \phi_1&=\textrm{atan}\left(\dot{\chi}\cot\chi/\dot{\varphi}\right)-\varphi,\\
\Omega &= -\dot{\chi}/\sin(\phi_1+\varphi),
\end{split}
\end{equation}
where the dot represents the time differential. To construct arbitrary dynamical gates, we divide the evolution path into two segments. At the first segment $[0,\tau/2]$, we set $\chi = \pi\sin^2(\pi t/\tau),f =\chi -\frac{1}{2}\sin(2\chi),$ and $\varphi = -\frac{2}{3}\sin^3\chi$. The resulting evolution operator is $U_D(\tau/2,0) = |d\rangle\langle d|+ e^{-i\pi/2}|e\rangle\langle b| + e^{i\pi/2}|b\rangle\langle e|$.  At the second segment $[\tau/2,\tau]$, we set $\chi = \pi\sin^2(\pi t/\tau),f =-\chi +\frac{1}{2}\sin(2\chi),$ and $\varphi = \frac{2}{3}\sin^3\chi-\gamma^\prime$. The resulting evolution operator is $U_D(\tau,\tau/2) = |d\rangle\langle d|+ e^{-i(\pi/2+\gamma^\prime)}|b\rangle\langle e| + e^{i(\pi/2+\gamma^\prime)}|e\rangle\langle b|$. Then, the total dynamical evolution operator can be expressed as $U_D(\tau,0) = |d\rangle\langle d|+ e^{-i(\pi+\gamma^\prime)}|b\rangle\langle b| + e^{i(\pi+\gamma^\prime)}|e\rangle\langle e|$; thus, arbitrary dynamical gates are constructed. In our experiment, the total evolution time $\tau = 105$~ns is chosen to ensure the same maximum coupling strength as SR-NHQC and NHQC gates. Note that, in general, the construction scheme of dynamical gates does not satisfy the condition (1) in the main text, and so does not exhibit robustness.

Now, we show how to measure dynamical phases $D_{mn}=\int_{0}^{\tau}d_{mn}(t) d t$ $(m, n=0, 1, 2)$ in our experiment. Here, we first measure the dynamical phase accumulated rate $d_{mn}(t) = \langle \psi_{m}(t)| H(t) |\psi_{n}(t) \rangle$ with $|\psi_{0}(t) \rangle = U(t,0)|d\rangle = |d\rangle$, $|\psi_{1}(t) \rangle = U(t,0)|b\rangle $, and $|\psi_{2}(t) \rangle = U(t,0)|e\rangle$. Since the dark sate is always decoupled, $d_{m0}(t)$ and $d_{0n}(t)$ are always zero at each time. Thus, we only need to measure $d_{mn}(t)$ for $m, n=1, 2$ at each evolution time, and then $D_{mn}$ can be obtained by an integration.

In order to measure $d_{11}(t)$ for a specific gate $U_1 (\theta, \phi, \gamma)$, we first prepare the initial state $| b\rangle = \sin(\frac{\theta}{2})e^{i\phi }| g\rangle + \cos(\frac{\theta}{2})|f\rangle$. Then, we apply a quantum gate $U_1 (\theta, \phi, \gamma)$ with confirmed Hamiltonian $H(t)$ to drive the qubit as $|\psi_{1}(t) \rangle = U(t,0)|b\rangle $. At each evolution time, we use the quantum state tomography technique to record the density matrix $\rho_{11}(t) = |\psi_{1}(t) \rangle\langle \psi_{1}(t)|$. Thus, $ d_{11}(t) = \langle \psi_{1}(t)| H(t) |\psi_{1}(t) \rangle = \textrm{Tr}[\rho_{11}(t)H(t)]$ can be obtained. Similarly, we can measure $d_{22}(t)$ for a specific gate $U_1 (\theta, \phi, \gamma)$ by preparing the initial state $| e\rangle$ and record the density matrix $\rho_{22}(t)  = |\psi_{2}(t) \rangle\langle \psi_{2}(t)|$ at each evolution time. Thus, $ d_{22}(t) = \langle \psi_{2}(t)| H(t) |\psi_{2}(t) \rangle = \textrm{Tr}[\rho_{22}(t)H(t)]$ can be obtained.

However, $d_{12}(t) = d^\dagger_{21}(t)$ cannot be directly measured. Here, we measure both its real and imaginary parts. In order to measure the real part $\textrm{Re}[d_{12}(t)]$, we first prepare the initial state $(| b\rangle + |e\rangle)/\sqrt{2}$ and apply a quantum gate $U_1(\theta, \phi, \gamma)$ with confirmed Hamiltonian $H(t)$ to drive the qubit. Then, we use the quantum state tomography technique at each evolution time to obtain the density matrix $\rho_{a1}(t)  = [|\psi_{1}(t) \rangle\langle \psi_{2}(t)|+|\psi_{2}(t) \rangle\langle \psi_{1}(t)|+|\psi_{1}(t) \rangle\langle \psi_{1}(t)|+|\psi_{2}(t) \rangle\langle \psi_{2}(t)|]/2 = [\rho_{12}(t)+\rho_{21}(t)+\rho_{11}(t)+\rho_{22}(t)]/2$. Thus, the real part of $d_{12}(t)$ can be obtained as 
\begin{eqnarray}
\textrm{Re}[d_{12}(t)] &=& \textrm{Re}[\langle \psi_{1}(t)| H(t) |\psi_{2}(t) \rangle] \notag\\
&=& \frac{1}{2} [\langle \psi_{1}(t)| H(t) |\psi_{2}(t) \rangle+\langle \psi_{2}(t)| H(t) |\psi_{1}(t) \rangle ] \notag\\
&=& \frac{1}{2}\textrm{Tr}[\rho_{21}(t)H(t) + \rho_{12}(t)H(t)] \notag\\
&=& \textrm{Tr}[\rho_{a1}(t)H(t)] - \frac{d_{11}(t)}{2} - \frac{d_{22}(t)}{2}. \notag
\end{eqnarray}
Similarly, we can measure the imaginary part $\textrm{Im}[d_{12}(t)]$ for a specific gate $U_1 (\theta, \phi, \gamma)$ by preparing the initial state $(| b\rangle -i |e\rangle)/\sqrt{2}$ and obtaining the density matrix $\rho_{a2}(t)  = [i|\psi_{1}(t) \rangle\langle \psi_{2}(t)|-i|\psi_{2}(t) \rangle\langle \psi_{1}(t)|+|\psi_{1}(t) \rangle\langle \psi_{1}(t)|+|\psi_{2}(t) \rangle\langle \psi_{2}(t)|]/2 = [i\rho_{12}(t)-i\rho_{21}(t)+\rho_{11}(t)+\rho_{22}(t)]/2$. Thus, the imaginary part can be obtained as 
\begin{eqnarray}
\textrm{Im}[d_{12}(t)] &=& \textrm{Im}[\langle \psi_{1}(t)| H(t) |\psi_{2}(t) \rangle] \notag\\
&=& \frac{1}{2}[-i\langle \psi_{1}(t)| H(t) |\psi_{2}(t) \rangle+i\langle \psi_{2}(t)| H(t) |\psi_{1}(t) \rangle ] \notag\\ 
&=& \frac{1}{2}\textrm{Tr}[-i\rho_{21}(t)H(t) + i \rho_{12}(t)H(t)] \notag\\
&=& \textrm{Tr}[\rho_{a2}(t)H(t)] - \frac{d_{11}(t)}{2} - \frac{d_{22}(t)}{2}. \notag
\end{eqnarray}

We have therefore measured the dynamical phase accumulated rate $d_{mn}(t)$ $(m, n=1, 2)$ at each evolution time. Thus, the total dynamical phases $D_{mn}$ can be obtained by a time integration of the corresponding rate $d_{mn}(t)$. We present the measured dynamical phase accumulated rate $d_{mn}(t)$ $(m, n=1, 2)$ for both the conventional NHQC and dynamical gates, as well as the SR-NHQC gate, in the main text.

\subsection{Analytical calculation of the robustness of SR NHQC}\label{B5}
Here, we theoretically calculate the gate fidelity as a function of the Rabi error $\varepsilon$ of the control field. Specifically, we consider the driving amplitude $\Omega(t)$ with an additional small error fraction of $\varepsilon\Omega(t)$. In other words, the Hamiltonian $H(t)$ becomes $H^{\prime}(t)=(1+\varepsilon) H(t)$. The corresponding evolution operator in the basis $\{|g\rangle, |e\rangle, |f\rangle\}$ is given by
\begin{equation}\label{Noise}
U_{\epsilon}(\tau)=\mathcal{T} e^{-i \int_{0}^{\tau} (1+\varepsilon) H(t) d t} \ .
\end{equation}
Applying perturbation theory~\cite{BLANES2009151} to Eq.~(\ref{Noise}), we have
\begin{equation}\label{Magus}
U_{\epsilon}(\tau,0)=I+\sum^{\infty}_{n=1}R_{n}(\tau,0) \ .
\end{equation}
The time-ordered products $R_{n}(\tau,0)$ are given by
\begin{equation}\label{product}
\begin{array}{l}
R_{1}(\tau,0)=(1+\varepsilon) \int_{0}^{\tau} d t_{1} {\bf d}(t_{1})\\
R_{2}(\tau,0)=(1+\varepsilon)^{2} \int_{0}^{\tau} d t_{1} \int_{0}^{t_{1}} d t_{2} {\bf d}(t_{1}) {\bf d}(t_{2}) \\
\vdots \\
R_{n}(\tau,0)=(1+\varepsilon)^{n} \int_{0}^{\tau} d t_{1} \ldots \int_{0}^{t_{n-1}} d t_{n} \left[{\bf d}(t_{1}) {\bf d}(t_{2}) \ldots {\bf d}(t_{n})\right] \ ,
\end{array}
\end{equation}
where ${\bf d(t)}=\sum_{m,n}d_{mn}(t)|\psi_{m}(0)\rangle\langle\psi_{n}(0)|$ is the instantaneous dynamical part. Applying the settings $d_{mn}=\langle\psi_{m}(t)|H(t)|\psi_{n}(t)\rangle$, $|\psi_{1(2)}(t)\rangle=U(t,0)|b(e)\rangle$, and ${\bf D}(\tau)=\int_{0}^{\tau} {\bf d}(t) dt=0 $ in the main text, the evolution operator of Eq.~(\ref{Noise}) in the basis $\{|\xi_{0}\rangle,|\xi_{1}(0)\rangle\}$ can be expressed as
\begin{equation}\label{NOIS}
U_{\varepsilon}(\tau, 0)=X\left|\xi_{1}(0)\right\rangle\left\langle\xi_{1}(0)|+| \xi_{0}\right\rangle\left\langle\xi_{0}\right| \ ,
\end{equation}
where $X\equiv \{1-\left(1-e^{i \gamma}\right) \cos ^{2} (\pi \varepsilon/2)\left[1+\sin ^{2} (\pi \varepsilon/2)\right] \}$.
Hence, a SR-NHQC gate under the Rabi control error in the basis $\{|g\rangle,|f\rangle\}$ is given by
\begin{equation}\label{NOISE01}
U_{1}^{\prime}(\theta, \phi, \gamma)=\left[\begin{array}{cc}
c_{\theta / 2}^{2}+s_{\theta / 2}^{2} X & \frac{(1-X)}{2} \mathrm{~s}_{\theta} e^{i \phi} \\
\frac{(1-X)}{2} \mathrm{~s}_{\theta} e^{-i \phi} & s_{\theta / 2}^{2}+c_{\theta / 2}^{2} X
\end{array}\right] \ ,
\end{equation}
where $c_{q}\equiv\cos(q/2)$, and $s_{q}\equiv\sin(q/2)$.
Using Eq.~(\ref{NOISE01}), the gate fidelity is
\begin{equation}\label{GateFidelity}
\begin{split}
\mathrm{F}(\varepsilon)&=\frac{1}{2}\left|\operatorname{Tr}\left[U_{1}^{\prime}(\theta, \phi, \gamma) U_{1}^{\dagger}(\theta, \phi, \gamma)\right]\right| \\
 &=\sqrt{\cos ^{2} \frac{\gamma}{2}+\sin ^{2} \frac{\gamma}{2} \cos ^{4} \frac{\pi \varepsilon}{2}\left(1+\sin ^{2} \frac{\pi \varepsilon}{2}\right)^{2}},
\end{split}
\end{equation}
where $U_{1}(\theta, \phi, \gamma)$ is the ideal evolution operator with $\epsilon=0$. When the error fraction $|\varepsilon|\ll 1$, $F(\varepsilon) \approx 1-\pi^{4}\varepsilon^4(1-\cos\gamma)/32$, exhibiting a  \emph{fourth}-order dependence on the Rabi error. 

\begin{figure}[tb]
	\centering
	\includegraphics{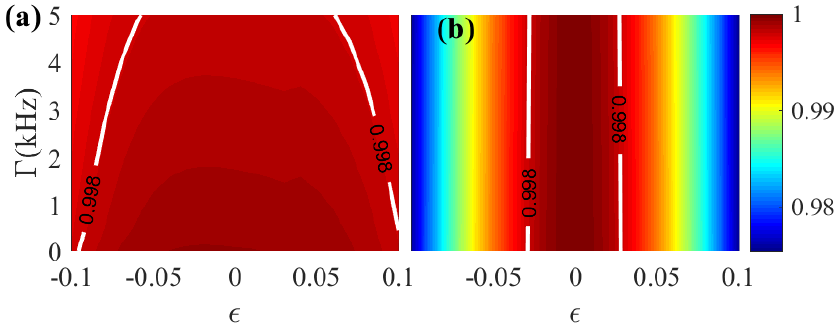}
	\caption{Numerical simulations of $X$ gate robustness via the SR-NHQC scheme (a) and conventional DRAG-based method (b) in the presence of both Rabi error with $\epsilon \in[-0.1,0.1]$ and  decoherence error with $\Gamma \in [0,5]$ kHz. The colorbar indicates the fidelity of the $X$ gate with dimensionless unit.}
	\label{Figs8}
\end{figure}

\subsection{Robustness of SR-NHQC gates against decoherence and Rabi errors}\label{B6}
The gate time of the demonstrated single-qubit SR-NHQC gates is generally longer than conventional DRAG-based gates, thus meaning that the gate infidelity mainly arises from a finite qubit coherence time. The ac Stark shift due to off-resonant levels in the weakly anharmonic qubits prevents us from achieving faster gate operations. The current results represent a balance between the efforts of reducing errors due to leakage and decoherence. With further improved qubit coherence time, the drawback associated with long gate operation times will be more and more compensated by the gained robustness. Here, we perform a comparison between SR-NHQC gates and conventional DRAG-based gates by numerically simulating the gate fidelity as a function of the Rabi error and qubit decoherence error rate, with the results shown in Fig.~\ref{Figs8}. Obviously, with improved qubit coherence times, coherence-limited errors for both methods are negligible. Since the robustness against the Rabi error for the SR-NHQC gates (fourth order) is obviously superior to the conventional DRAG scheme (only second order), our SR-NHQC approach is more preferable in the near future.

\section{TWO-QUBIT SR-NHQC GATES}\label{C}
\subsection{Initialization of two-qubit states}\label{C1}
\begin{figure}[tb]
	\centering
	\includegraphics{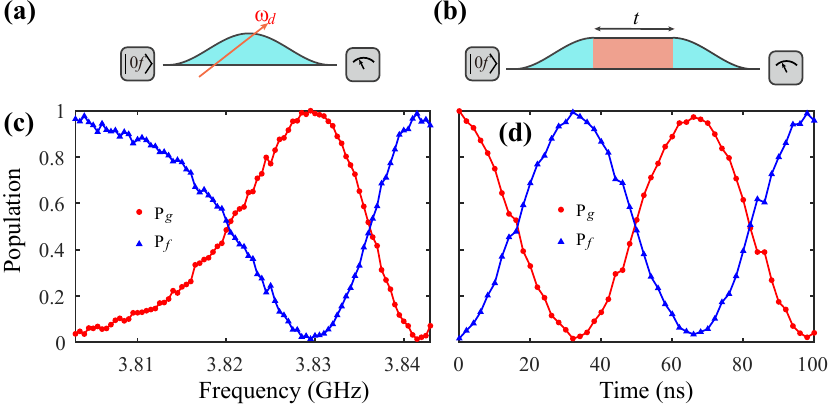}
	\caption{Characterization of the Raman transition drives. (a) Pulse sequence to find the resonant frequency of the Raman transition drive to achieve the $\ket{0f} \leftrightarrow \ket{1g}$ coupling. The drive pulse has a cosine-shape envelope with a duration of 140~ns. (b) Pulse sequence to perform the Rabi oscillation between the $\ket{0f}$ and $\ket{1g}$ states. (c) Measured qubit populations $\textrm{P}_g$ and $\textrm{P}_f$ as a function of the driving frequency $\omega_d$, from which we confirm the resonant driving frequency $\omega_d = 3.83$ GHz. (d) Measured qubit populations $\textrm{P}_g$ and $\textrm{P}_f$ as a function of the plateau of the drive pulse.}
	\label{Figs9}
\end{figure}

Before implementing the two-qubit SR-NHQC gates, initialization of two-qubit states in the two-qubit subspace \{$\ket{0g}$, $\ket{0f}$, $\ket{2g}$, $\ket{2f}$\} is needed. Qubit initial states can be easily prepared by sequential $\pi$ pulses on $\ket{g}\leftrightarrow \ket{e}$ and $\ket{e}\leftrightarrow \ket{f}$ transitions, while the initial Fock state of the cavity is more difficult. Here, we present the details of initialization of the cavity states, which is generated through sequential Raman transition drives on $\ket{0f}\leftrightarrow\ket{1g}$ and $\ket{1f}\leftrightarrow\ket{2g}$, respectively. The Raman transition drive of $\ket{0f}\leftrightarrow\ket{1g}$ is calibrated by first preparing initial state $\ket{0f}$ through qubit sequential $\pi$ pulses on $\ket{g}\leftrightarrow\ket{e}$ and $\ket{e}\leftrightarrow\ket{f}$ transitions, then applying a cosine-shaped microwave pulse with a variable driving frequency to achieve the $\ket{0f} \leftrightarrow \ket{1g}$ coupling, and finally measuring the qubit populations. The corresponding pulse sequence and experimental results are shown in Figs.~\ref{Figs9}(a,c). Thus, we could extract the resonant frequency of the Raman transition drive to achieve the $\ket{0f} \leftrightarrow \ket{1g}$ coupling. Besides, we further measure the Rabi oscillation between $\ket{0f}$ and $\ket{1g}$ states, with the corresponding pulse sequence and measurement results shown in Figs.~\ref{Figs9}(b,d). After performing similar procedures, we could also calibrate the Raman transition drive between $\ket{1f}$ and $\ket{2g}$ states. Therefore, initial Fock states $\ket{2g}$ and $\left(\ket{0g} + \ket{2g} \right)/\sqrt{2}$ can be generated by the evolution sequence $ \ket{0g}\rightarrow\ket{0e} \rightarrow\ket{0f} \rightarrow\ket{1g} \rightarrow\ket{1e} \rightarrow\ket{1f} \rightarrow\ket{2g}$ and $\ket{0g}\rightarrow\ket{0e} \rightarrow (\ket{0e}+\ket{0f})/\sqrt{2} \rightarrow(\ket{0e}+\ket{1g})/\sqrt{2} \rightarrow(\ket{0g}+\ket{1e})/\sqrt{2} \rightarrow(\ket{0g}+\ket{1f})/\sqrt{2} \rightarrow(\ket{0g}+\ket{2g})/\sqrt{2}$, respectively, where each evolution step is realized by a $\pi$ or $\pi/2$ rotation on the corresponding transitions. The generated initial Fock states are calibrated through a number splitting experiment, with the measurement results shown in Fig.~4(b) in the main text.

\subsection{State population evolution of two-qubit SR-NHQC gates}\label{C2}
\begin{figure}[t]
	\centering
	\includegraphics{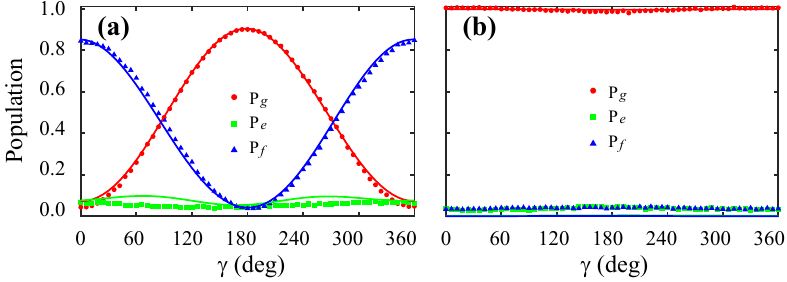}
	\caption{State population of nontrivial two-qubit SR-NHQC gates $U_2(\pi/2,0, \gamma)$ as a function of $\gamma$ for initial states $|0f\rangle$ (a) and $|2g\rangle$ (b), respectively, consistent with the numerical simulations~(solid lines).  }
	\label{Figs10}
\end{figure}

The two-qubit SR-NHQC gates are implemented by applying two microwave drives resonantly with the $\ket{0g}\leftrightarrow \ket{0e}$ and $\ket{0e}\leftrightarrow\ket{0f}$ transitions, while keeping $\ket{2g}$ and $\ket{2f}$ unaffected thanks to the strong dispersive $ZZ$ interaction. The total evolution time $\tau = 2760$~ns is chosen in our experiment such that the two drives are weaker enough to avoid driving undesired transitions. As for the single-qubit SR-NHQC gate, the total evolution time is divided into six segments with the driving parameters also satisfying Eq.~(\ref{splitpulses}). In the main text, we have demonstrated the arbitrary tunability of the parameter $\gamma$ for the two-qubit SR-NHQC gates $U_2(\pi/2,0,\gamma)$ for an initial state $\left(|0g\rangle + |2g\rangle\right)/\sqrt{2}$. Here, we present extended data in Fig.~\ref{Figs10}, where we have measured the qubit state populations as a function of $\gamma$ for the two-qubit SR-NHQC gate $U_2(\pi/2,0,\gamma)$ with initial states $|0f\rangle$  and $|2g\rangle$, respectively. For the two-qubit SR-NHQC CNOT gate defined by $U_2(\pi/2,0, \pi)$, a gate fidelity of 0.944 is obtained from the average of the correct state populations of the CNOT gate operation on initial states $\ket{0f}$ and $\ket{2g}$, respectively. The infidelity mainly comes from the decoherence of the qubit due to a long gate operation time and can be reduced with a short gate time when implemented on a system of two transmon qubits and a tunable coupler.

\section{GATE ERROR ANALYSIS}\label{D}
In this section, we estimate the main error sources and their contributions to the loss of fidelity for both the single- and two-qubit SR-NHQC gates.

\begin{table*}[htb]
\caption{Infidelities of the single- and two-qubit SR-NHQC gates.}
\begin{tabular}{p{5cm}<{\centering} p{4.6cm}<{\centering} p{4.6cm}<{\centering} }
  \hline
  Error sources & Single-qubit SR-NHQC gates & Two-qubit SR-NHQC gates\\\hline
  &&\\[-0.9em]
  Decoherence error of transmon qubit & 0.0043 & 0.050\\
  &&\\[-0.9em]
  Decoherence error of photonic qubit & - & 0.008\\
  &&\\[-0.9em]
  Driving-induced leakage error & 0.0006 & $<$ 0.001  \\
  &&\\[-0.9em]
  Total & 0.0049 & 0.058  \\
  &&\\[-0.9em]
  \hline
\end{tabular}
\label{TableS3}
\end{table*}

1. The decoherence errors come from the qubits relaxation and dephasing processes during the gate operation. For the single-qubit SR-NHQC gates, we estimate the decoherences errors with the equation
\begin{equation}\label{decayerror}
e_c \approx \frac{1}{9}(2\Gamma_1+2\Gamma_2+2\Gamma_3+\Gamma_{ge}+\Gamma_{ef})\tau,
\end{equation}
according to Ref.\cite{Morvan2021Qutrit}, where $\tau$ is the gate time; $\Gamma_{ge}$ and $\Gamma_{ef}$ are the corresponding relaxation rates; $\Gamma_{1} = 1/T^E_{2_{ge}}$, $\Gamma_{2} = 1/T^E_{2_{ef}}$, and $\Gamma_{3} = 1/T^E_{2_{gf}}$ are the corresponding dephasing rates of the three energy levels with $T_{2^E_{ge}}$, $T_{2^E_{ef}}$, and $T^E_{2_{gf}}$ the qubit dephasing times extracted from echo experiments. The time $T_{2^E_{ge}} = 38 \mu$s is directly measured from the corresponding experiment. Because of the lack of direct measurements of $T_{2^E_{ef}}$ and $T_{2^E_{gf}}$, we estimate these two dephasing times as $T_{2^E_{ef}} = 2/\Gamma_{ef}\approx26\mu$s and $T_{2^E_{gf}} = 2/(\Gamma_1 + \Gamma_2)\approx31\mu$s, respectively. For the single-qubit SR-NHQC gates with a gate time of $120$~ns, the estimated coherence-limited error is about $4.3\times 10^{-3}$. For the two-qubit SR-NHQC gates with a gate duration of $2760$~ns, the decoherence error is estimated by the average errors from the decoherence of both the transmon qubit and photonic qubit, resulting in a decoherence error of $5.8\times 10^{-2}$.

2. There are also small contributions from the driving-induced leakage errors due to the weakly anharmonic qubits. We estimate these leakage errors from numerical simulations without considering any decoherence of the qubits. The simulation results give leakage errors of $6\times 10^{-4}$ and $<1\times 10^{-3}$ for the single- and two-qubit SR-NHQC gates, respectively.

We summarized these results in Table.~\ref{TableS3}. As a result, the totally estimated infidelities are consistent with our experimental results for both the single- and two-qubit SR-NHQC gates and the dominant error source is the qubits decoherence for both gates. 

%

\end{document}